\documentclass[%
reprint,
superscriptaddress,
amsmath,amssymb,
aps,
pra,
prstab,
]{revtex4-2}
\usepackage{xcolor}
\usepackage{graphicx}
\usepackage{dcolumn}
\usepackage{bm}
\usepackage{booktabs}
\usepackage{multirow}
\usepackage{amssymb}
\usepackage{float}
\usepackage{epstopdf}

\begin{document}

\preprint{pentagon}
\title{Wave functions for the regular pentagonal two-dimensional quantum box and thin  microstrip antenna}

\author{Tristan Langhorne}
\affiliation{Department of Physics, University of Central Florida, Orlando, FL 32816-2385, United States of America}
\author{Erik E. Domenech}
\affiliation{Department of Physics, University of Central Florida, Orlando, FL 32816-2385, United States of America}
\author{Juan Oliveros Gonzalez}
\affiliation{Department of Physics, University of Central Florida, Orlando, FL 32816-2385, United States of America}
\author{Richard A. Klemm}
\email{richard.klemm@ucf.edu}
\affiliation{Department of Physics, University of Central Florida, Orlando, FL 32816-2385, United States of America}


\date{\today}
\begin{abstract} The general wave functions for the two-dimensional regular pentagonal quantum box and thin microstrip antenna are derived.  As for the square, equilateral triangular, and circular disk-shaped boxes and antennas, there are two quantum nunbers $n$ and $m$.  In those cases, $n\ge1 $ and $m\ge 0$ are both unlimited non-negative integers of any value. For the regular pentagon, only $n\ge1 $ is an unlimited positive quantum number, but $m_{\rm min}\le m\le 5$, where $m_{\rm min}=0$ for the pentagonal microstrip antenna with Neumann boundary conditions and $m_{\rm min}=1$ for the pentagonal quantum box with Dirichlet boundary conditions.  Color-coded pictures of the wave functions for the regular pentagonal quantum box and microstrip antenna are presented for all allowed $m$ values and for $1\le n\le 2$ and for the microstrip antenna for all allowed $m$ values and $n=3$.

\end{abstract}

\maketitle

\section{Introduction}
There has been a considerable interest in two-dimensional quantum boxes and thin microstrip antennas of a variety of shapes [1-49].  The most common shape studied experimentally has been the rectangle [1,5-12,15-17,19-22,25-33,35,37,38,41-47,49] or arrays of rectangular mesas in order to increase the output power of THz emission from the intrinsic Josephson junctions in the high-temperature superconductor Bi$_2$Sr$_2$CaCu$_2$O$_{8+\delta}$ (Bi2212)  [10,14,15,18,40].  But there have also been theoretical and/or experimental studies on squares [3,4,7,17,30,34,35,46], truncated squares [45], equilateral triangles [2,4,17,50], acute isosceles triangles [11,14,17,48], right isosceles triangles [17], cylindrical or circular disks [4,6,8,11,14,17,25,30], ellipses [37], cross whiskers [40], annuli [49], slitted annuli [49], arrows [44], and one experimental study of a regular pentagonal mesa [24].
  
We  first describe a regular pentagon as being inscribed inside a circle of radius $\alpha$ with one corner at the point $A=\alpha(-1,0)$ and the other corners are at 
$B=\alpha[-\sin(\pi/10),\cos(\pi/10)]$, $C=\alpha[\cos(\pi/5),\sin(\pi/5)]$,  $D=\alpha[\cos(\pi/5),-\sin(\pi/5)]$, and $E=\alpha[-\sin(\pi/10),-\cos(\pi/10)]$, as sketched in Fig. 1.  In this notation, the length $a$ of each side of the regular pentagon is that of the vertical line from $C$ to $D$, which is 
\begin{eqnarray}
a&=&2\alpha\sin(\pi/5)\nonumber\\
&=&\alpha\sqrt{\frac{5-\sqrt{5}}{2}}.
\end{eqnarray}
\begin{figure}
{\includegraphics[width=0.45\textwidth]{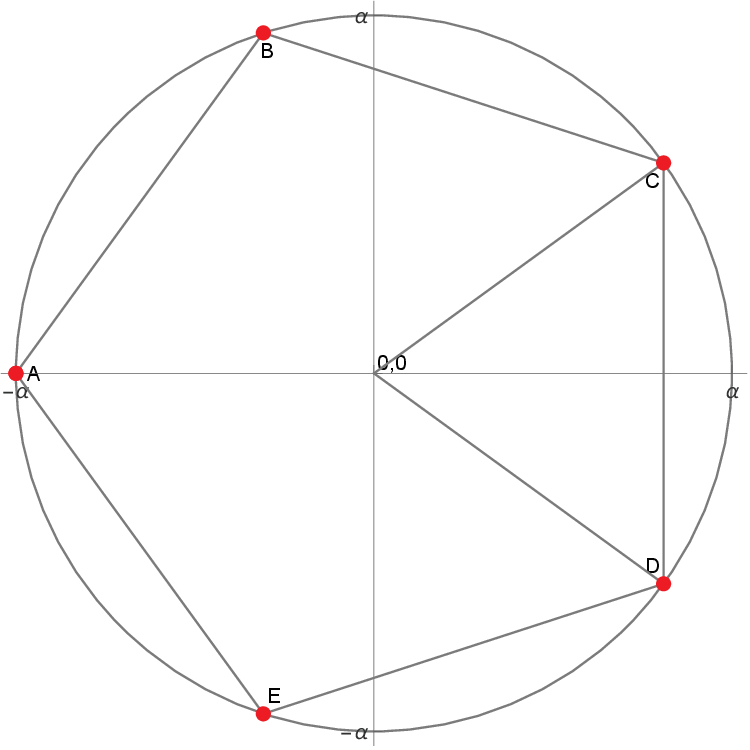} 
\caption{Plot of the boundary of a regular pentagon with sides of length $a$ inscribed inside a circle of radius $\alpha$ with its corners labelled $A, B, C, D, E$, where corner $A$ is the far left corner and the other four corners are labelled in a consecutive clockwise fashion as pictured.}
\label{fig1}}
\end{figure}

For both a two-dimensional regular pentagonal box and a thin regular pentagonal microstrip antenna, the  quantum wave function $\Psi(x,y)$ respectively for a particle or ''effective particle'' of mass $M$ satisfies the Schr{\"o}dinger equation,
\begin{eqnarray}
\Bigl[-\frac{\hbar^2}{2M}\Bigl(\frac{\partial^2}{\partial x^2}+\frac{\partial^2}{\partial y^2}\Bigr)+\Phi(x,y)\Bigr]\Psi(x,y)&=&E\Psi(x,y),
\end{eqnarray}
where the electrostatic potential  $\Phi(x,y)=0$ inside of and $\Phi(x,y)=\infty$ on and outside of the circumference of the quantum pentagonal box with the Dirichlet boundary condition.  For the microstrip antenna, we are mostly interested in thin pentagonal mesas of the high-temperature superconductor Bi$_2$Sr$_2$CaCu$_2$O$_{8+\delta}$ (Bi2212), for which one applies a current in the $z$ direction that generates an effective magnetic field $H_z(x,y)$, so that $\Phi(x,y)=0$ and the wave function $\Psi(x,y)\propto H_z(x,y)$, the planar normal derivative of which vanishes on the circumference of the pentagonal antenna with the Neumann boundary condition  [1]-[8].

As shown in detail in Appendix 1,  the exact wave functions for a particle within the portion of the quantum box inside the isosceles triangular region of Fig. 1 with corners at the origin of the circle and at points C and D in the region $0\le x\le\alpha\cos(\pi/5)$ that vanish (the Dirichlet condition) along the  vertical axis  at $x=\alpha\cos(\pi/5)$ that are even and odd about the horizonal axis are respectively
\begin{eqnarray}
\Psi^{(e)}_{nm}(x,y)&=&N\sin\Bigl(\frac{nm\pi x}{\alpha\cos(\pi/5)}\Bigr)\cos\Bigl(\frac{(5-m)n\pi y}{\alpha\sin(\pi/5)}\Bigr),\label{evenbox}\\
\Psi^{(o)}_{nm}(x,y)&=&N\sin\Bigl(\frac{nm\pi x}{\alpha\cos(\pi/5)}\Bigr)\sin\Bigl(\frac{(5-m)n\pi y}{\alpha\sin(\pi/5)}\Bigr).\label{oddbox}
\end{eqnarray}
The exact wave functions for an effective particle  within the same isosceles triangular region of the thin microstrip antenna, for which the normal derivatives with respect to $x$ evaluated at $x=\alpha\cos(\pi/5)$ vanish (the Neumann condition) are also shown in Appendix 1 to be
\begin{eqnarray}
\Psi^{(e)}_{nm}(x,y)&=&N\cos\Bigl(\frac{nm\pi x}{\alpha\cos(\pi/5)}\Bigr)\cos\Bigl(\frac{(5-m)n\pi y}{\alpha\sin(\pi/5)}\Bigr),\label{evenantenna}\\
\Psi^{(o)}_{nm}(x,y)&=&N\cos\Bigl(\frac{nm\pi x}{\alpha\cos(\pi/5)}\Bigr)\sin\Bigl(\frac{(5-m)n\pi y}{\alpha\sin(\pi/5)}\Bigr),\label{oddantenna}
\end{eqnarray}
where the normalization constant $N$ is shown in Appendix 2 to be
\begin{eqnarray}
N&=&\frac{4}{a}\sqrt{\frac{\tan(\pi/5)}{5}}\\
&=&\Bigl[\frac{2^8(\sqrt{5}-2)}{5^{3/2}}\Bigr]^{1/4}/a\\
&\approx&1.52477/a,
\end{eqnarray}
which is independent of the quantum numbers $n$ and $m$ and is the same for all four of these wave functions for this isosceles triangular region. The normalization constant $N$ was determined by setting the probability of finding the particle or effective particle  within the equilateral triangular portion of the pentagonal box or antenna to be $\frac{1}{5}$.  

For all four above general wave functions, $n\ge1$ is unbounded.
In general $0\le m\le 5$, but for the box and antenna wavefunctions odd about the horizontal axis, $m\le 4$, and for the box wave functions even about the horizontal axis, $1\le m\le 5$.  The energies for the particle or effective particle represented by both general wave functions are given by
\begin{eqnarray}
E_{n,m}&=&\frac{\hbar^2\pi^2n^2}{2M\alpha^2}\Biggl(\frac{m^2}{\cos^2(\pi/5)}+\frac{(5-m)^2}{\sin^2(\pi/5)}\Biggr),
\end{eqnarray}
where the mass $M$ is only relevant for the particle in the 2D pentagonal box, and has a different interpretation for the effective particle in the thin microstrip antenna. 

For a  thin microstrip antenna consisting of a material with index of refraction $n_r$, after activating the antenna such as by applying a voltage $V$ normal to the layers in Bi2212, the frequencies $f_{n,m}$ of emission are given by [1]-[8],
\begin{eqnarray} 
f_{n,m}&=&\frac{c_0}{2n_r}\Bigl[\frac{2ME_{n,m}}{\pi^2\hbar^2}\Bigr]^{1/2},
\end{eqnarray}
which for a thin regular pentagonal microstrip antenna become
\begin{eqnarray}
f_{n,m}&=&\frac{\sqrt{2}nc_0}{n_r\alpha}\Bigl[\frac{m^2}{3+\sqrt{5}}+\frac{(5-m)^2}{5-\sqrt{5}}\Bigr]^{1/2},
\end{eqnarray}
where $c_0$ is the speed of light in vacuum.
We note that $n_r\approx 4.2$ for Bi2212.  Thus, $n_r\alpha/c_0$, which has units of seconds, plays the role of the effective mass in the thin microstrip antenna [1]-[8].

As shown in the following, in order to construct the entire wave functions for the regular pentagonal quantum box and the regular pentagonal microstrip antenna, only those wave functions in Eqs. (3) and (5) that are even about the horizonal axis can be employed by rotations about the origin of the enclosed circle by multiples of $2\pi/5$ radians in order to obtain overall wave functions that are everywhere continuous inside the pentagon. 

That is, for the wave functions in Eqs. (4) and (6) for the isosceles triangular region pictured in Fig. 1, in order to guarantee the matching of that wave function with the version of it  rotated counterclockwise by $\pi/5$ radians, one needs to change the overall sign of the rotated isosceles triangle.  Then, the second rotated isosceles triangle needs to have the same sign as the original one, the third rotated isosceles triangle needs the same sign as the first  rotated one, and the fourth isosceles triangle would need to have the same rotated sign as the second one to match the boundary between the third and fourth rotated isosceles triangle.  But this fourth rotated isosceles triangle would then have a non-matching sign with the original isosceles triangle given by either Eq. (4) or (6).  Thus,  the wave functions in Eqs. (4) and (6) cannot be used to respectively construct the full wave functions for the two-dimensional pentagonal quantum box and microstrip antenna. However, this suggests that for the regular pentagonal antenna, cutting a slit from the center of the pentagon to one of its corners could increase  the output power by a factor of 4.  The wave functions in Eq. (6) for that  singly slitted thin regular pentagonal antenna will be presented in a subsequent paper. 

The wave functions for the two-dimensional regular pentagonal box with $n=1,2$ and all allowed $m$ values are presented in Sec. II.  The wave functions for the thin regular pentagonal microstrip antenna with $n=1,2$ and all allowed $m$ values are presented in Sec. III.  In Appendix 1 (Sec. IV),  the derivation  of the general form of the wave functions for the two-dimensional regular pentagonal box is presented.  In Appendix 2 (Sec. V), the normalization constant $N$ is derived for the box.  In Appendix 3 (Sec. VI), the Mathematica\copyright  code used to generate the figures is presented.
  In Appendix 4 (Sec. VII), the wave functions for $n=3,4$ and the allowed $m=1,2,3,4,5$ values for the two-dimensional regular quantum box are shown.  In Appendix 5 (Sec. VIII), the wave functions for $n=3$ and the allowed $m=0,1,2,3,4,5$ values for the thin regular pentagonal microstrip antenna are presented.  
The character table for the two-dimensional regular pentagonal group $C_{5v}$ is given in Table I [9]. The operations $C_5$ are rotations about the central axis by $\frac{1}{5}$ of a circle in either direction.  The operations $\sigma_v$ are mirror planes on both sides of the hexagonal plane.
\vskip5pt
\begin{table}
\caption{Regular pentagonal group $C_{5v}$ character table [9]}
\begin{center}
 \begin{tabular}{|| c | c | c | c | c ||} 
 \hline
  $C_{5v} (5m)$ & $E$ & $2C_5$ & $2C_5^2$ & $5\sigma_v$   \\
 \hline\hline
 $A_1$ & 1& 1& 1 & 1 \\ 
 \hline
 $A_2$ & 1 & 1 & 1 & -1\\
 \hline
 $E_1$& 2 & $2\cos(2\pi/5)$ & $2\cos(4\pi/5)$ & 0 \\
 \hline
 $E_2$ & 2 &$2\cos(4\pi/5)$ & $2\cos(8\pi/5)$& 0 \\
 \hline
\end{tabular}
\end{center}
\end{table}

\section{Plots of the wave functions with $n=1,2$ of the regular two-dimensional pentagonal box}
In order to protray the wave functions for the regular pentagonal box, the general form of which is given in Eq. (3), we used Mathematica to plot the isosceles triangular portion of box extending from the origin to the points at C and D in Fig. 1, and to rotate it four times by $2\pi/5$ about the origin. As noted above,  the wave functions in Eq. (4) are odd about the horizontal axis, and cannot be used to construct the full wave functions of the pentagonal box.  
For all of the allowed wave functions, the color-coded plots are normalized equivalently, and the values according to each of the colors is indicated by the bar code presented in Fig. 2.
\begin{figure}
{\includegraphics[width=0.10\textwidth]{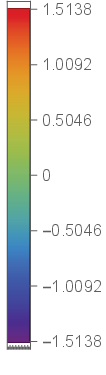} 
\caption{Bar code indicating the numerical values of the colors in all of the normalized wave function plots.}
\label{fig2}}
\end{figure}
Color plots of the wave functions for the regular pentagonal quantum box with $n=1$ and $m=1$ to 5  are shown in Figs. 3 - 7.  Color plots of the wave functions for the regular pentagonal quatum box  with $n=2$ and $m=1$ to 5 in are shown in Figs. 8 - 12.  In each of these figures, the wave functions are continuous everywhere inside the box and due to the Dirichlet boundary condition that the wave functions all vanish on the boundaries, the pentagonal boundaries are portrayed in black.  Note that each of these 10 figures has the property that there are $nm$ concentric black regular pentagons, at which the wave functions vanish.  It is also noteworthy that since the boundary contains discontinuous derivatives at the five corner points, the wave functions also exhibit some additional discontinuous derivatives along the lines from the center to the five corners of the pentagonal boxes.
\begin{figure}
{\includegraphics[width=0.45\textwidth]{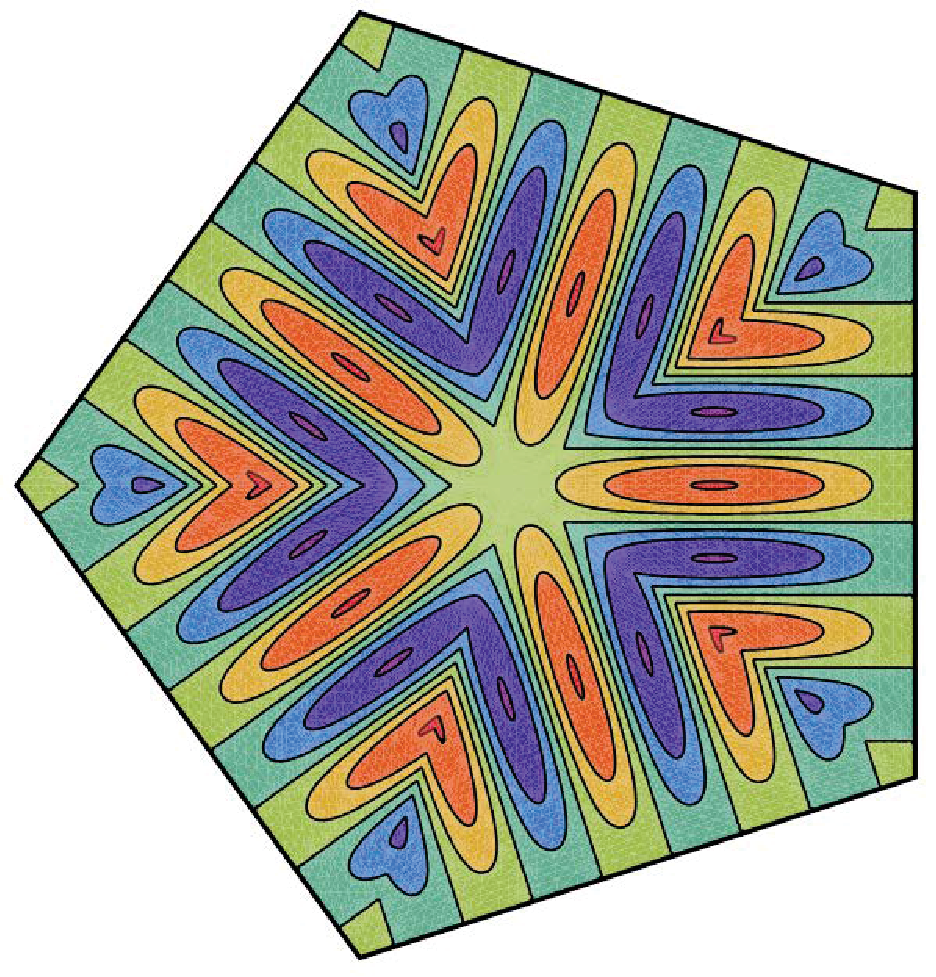} 
\caption{Plot of the normalized wave function from Eqs. (3) and (7) for the regular pentagonal quantum box with $m=1, n=1$.}
\label{fig3}}
\end{figure}
\begin{figure}
{\includegraphics[width=0.45\textwidth]{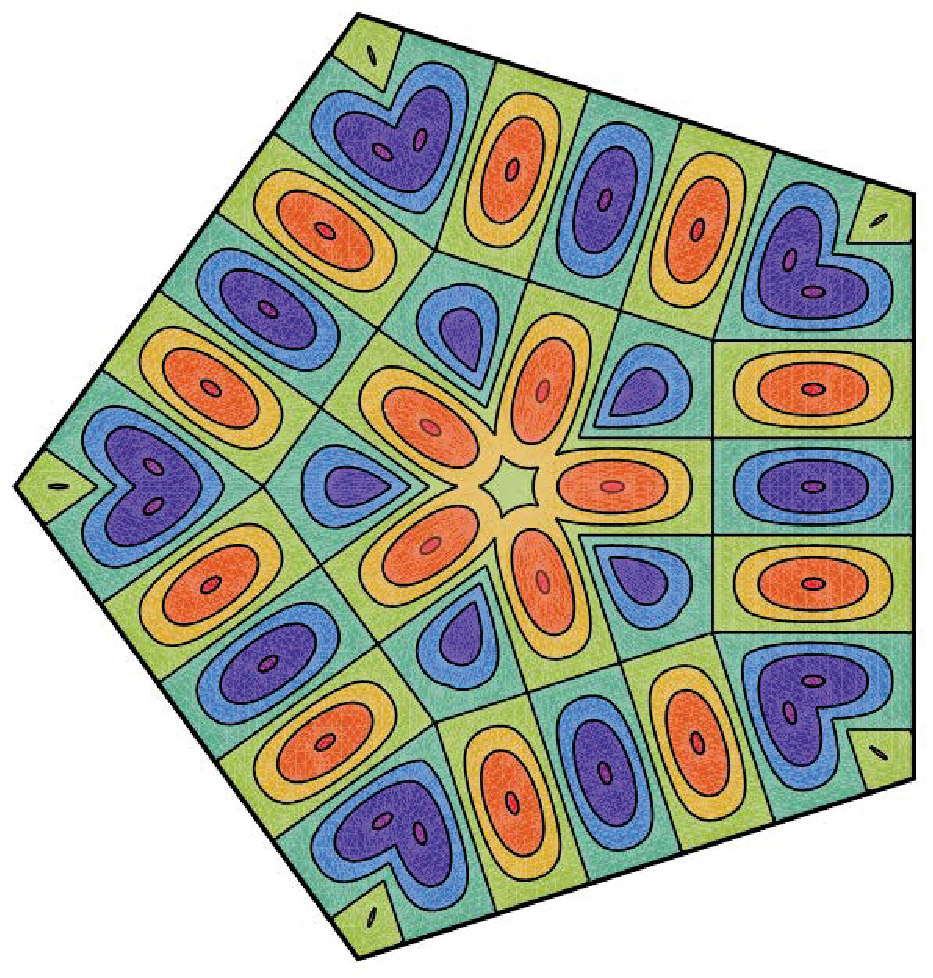} 
\caption{Plot of the normalized wave function from Eqs. (3) and (7) for the regular pentagonal quantum box with $m=2, n=1$. }
\label{fig4}}
\end{figure} 
\begin{figure}
{\includegraphics[width=0.45\textwidth]{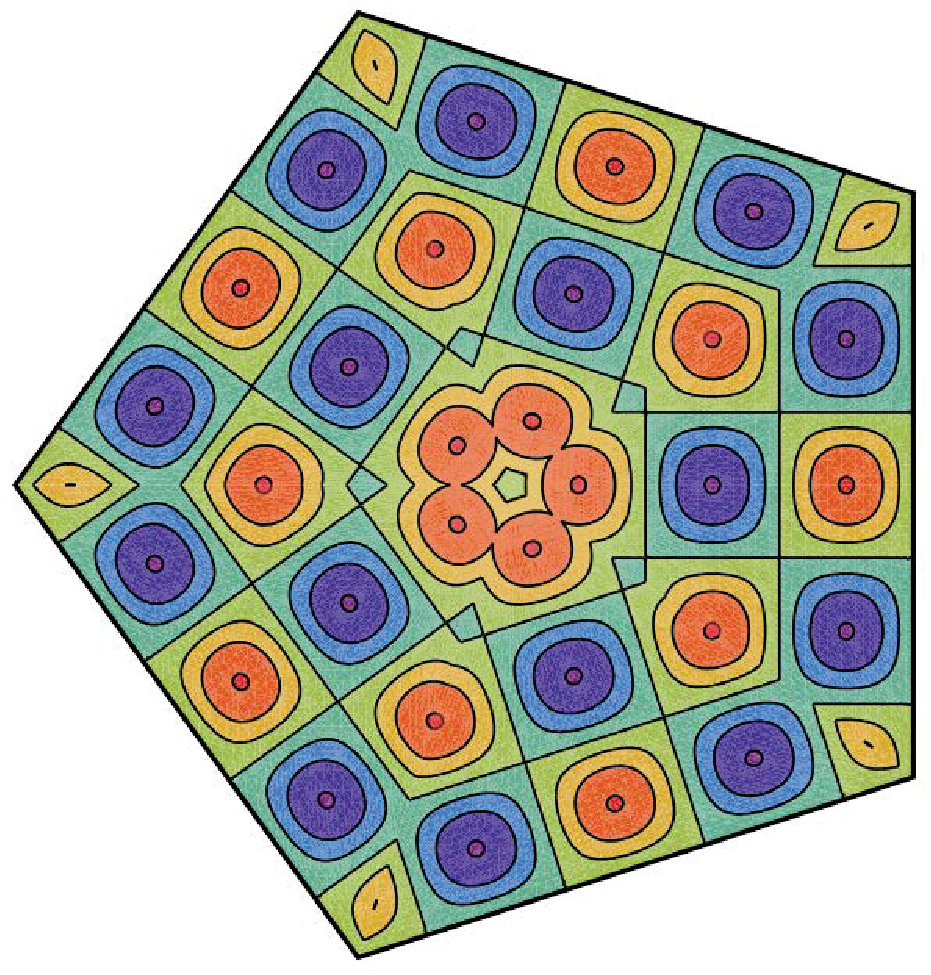} 
\caption{Plot of the normalized wave function from Eqs. (3) and (7) for the regular pentagonal quantum box with $m=3, n=1$. }
\label{fig5}}
\end{figure} 
\begin{figure}
{\includegraphics[width=0.45\textwidth]{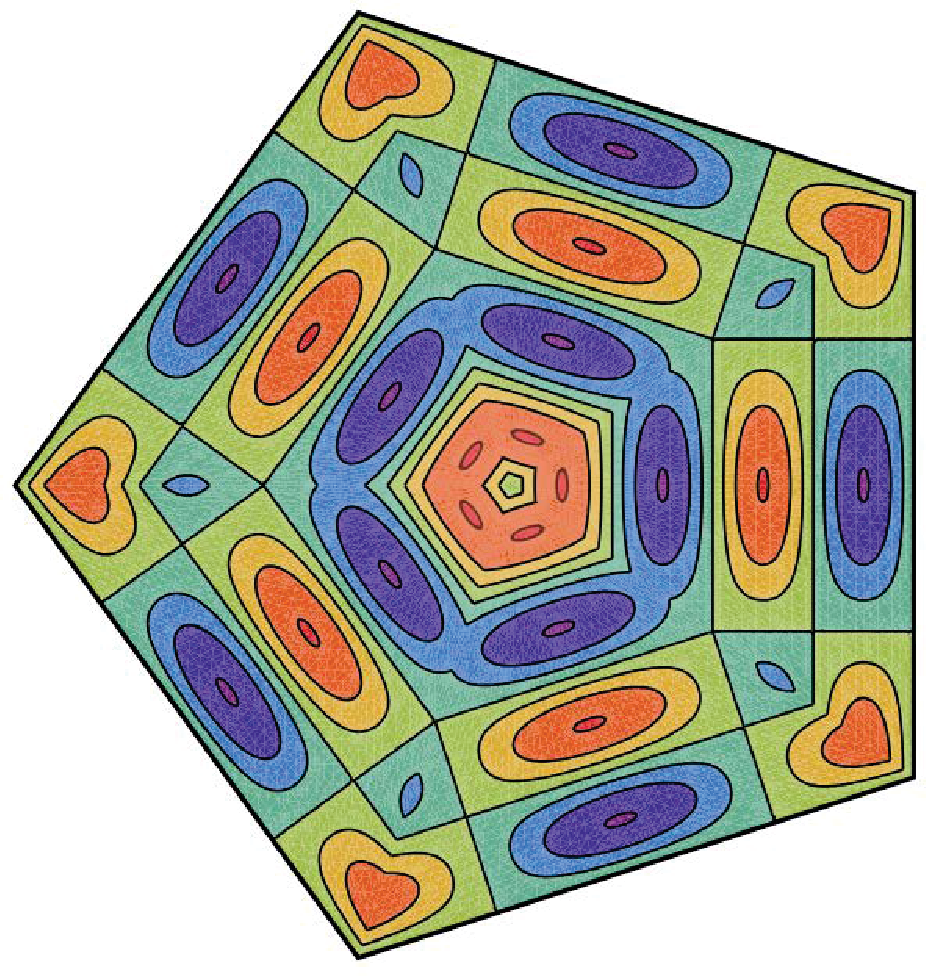} 
\caption{Plot of the normalized wave function from Eqs. (3) and (7) for the regular pentagonal quantum box with $m=4, n=1$. }
\label{fig6}}
\end{figure}
\begin{figure}
{\includegraphics[width=0.45\textwidth]{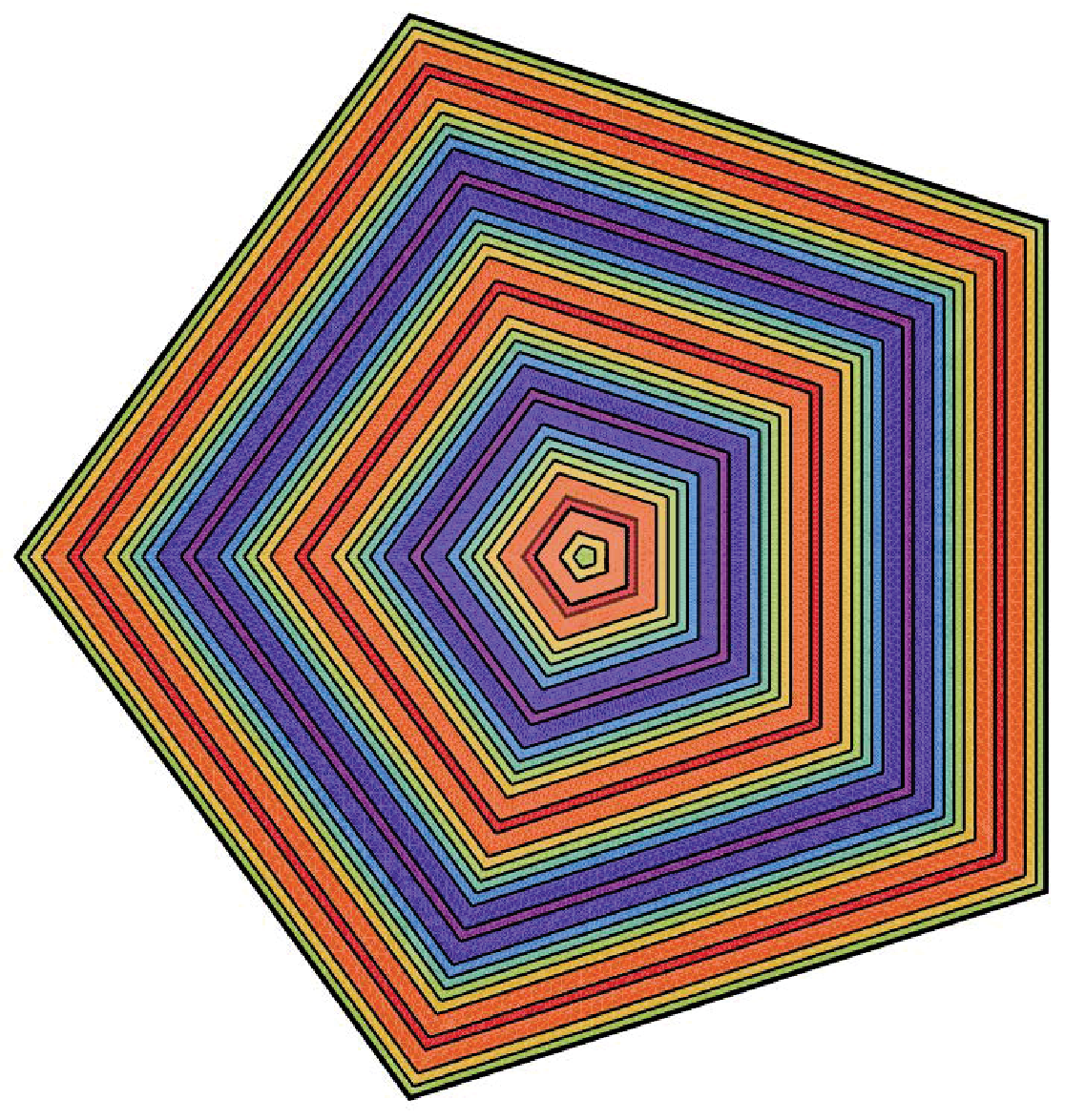} 
\caption{Plot of the normalized wave function from Eqs. (3) and (7) for the regular pentagonal quantum box with $m=5, n=1$. }
\label{fig7}}
\end{figure}
\begin{figure}
{\includegraphics[width=0.45\textwidth]{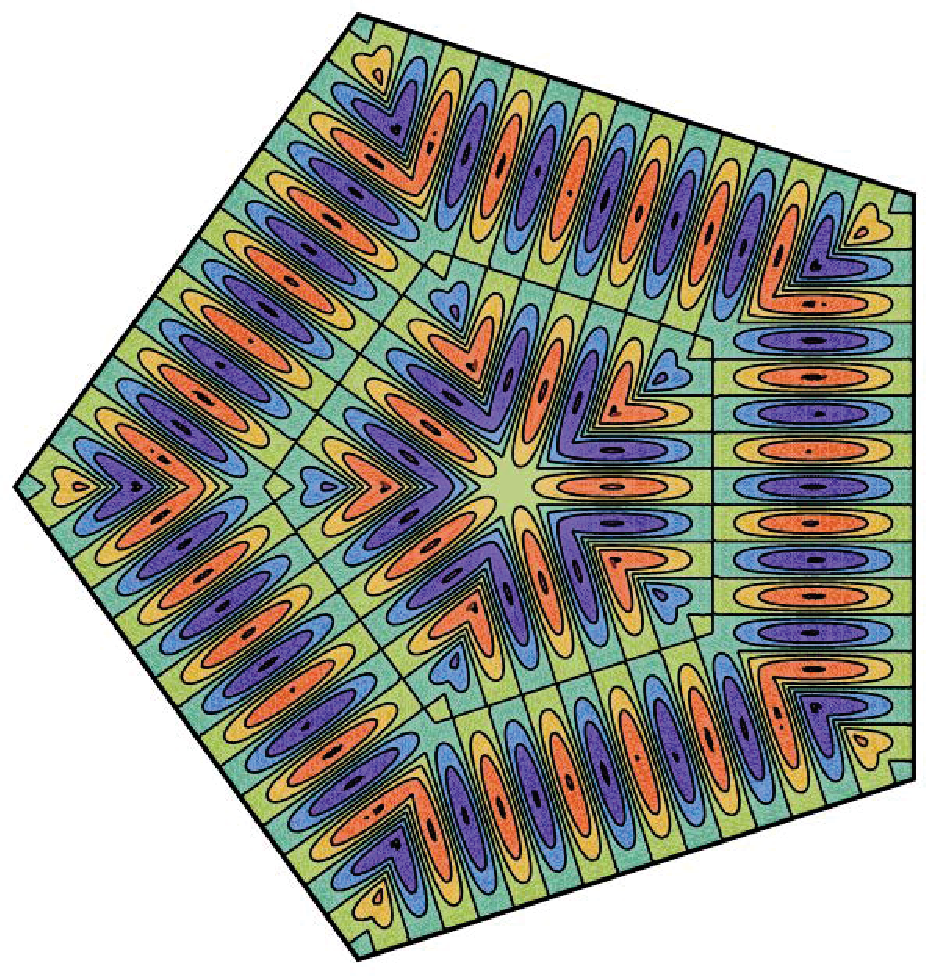} 
\caption{Plot of the normalized wave function from Eqs. (3) and (7) for the regular pentagonal quantum box with $m=1, n=2$. }
\label{fig8}}
\end{figure}
\begin{figure}
{\includegraphics[width=0.45\textwidth]{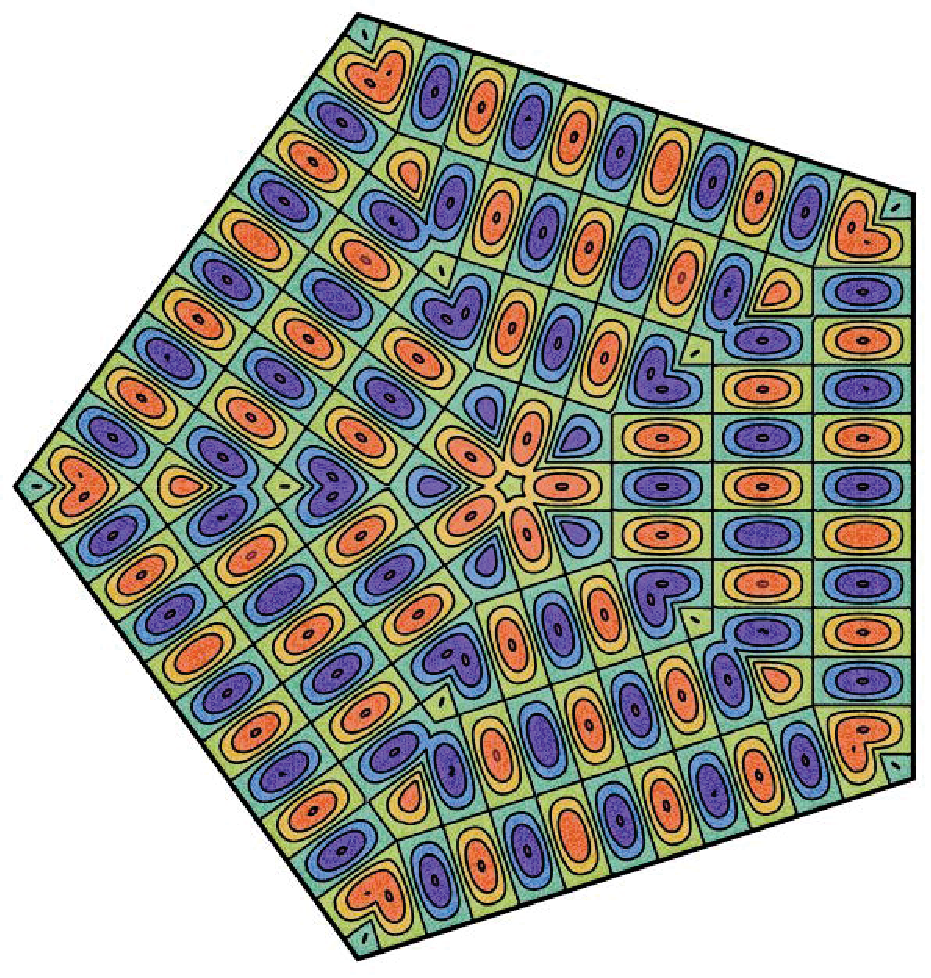} 
\caption{Plot of the normalized wave function from Eqs. (3) and (7) for the regular pentagonal quantum box with $m=2,n=2$. }
\label{fig9}}
\end{figure}
\begin{figure}
{\includegraphics[width=0.45\textwidth]{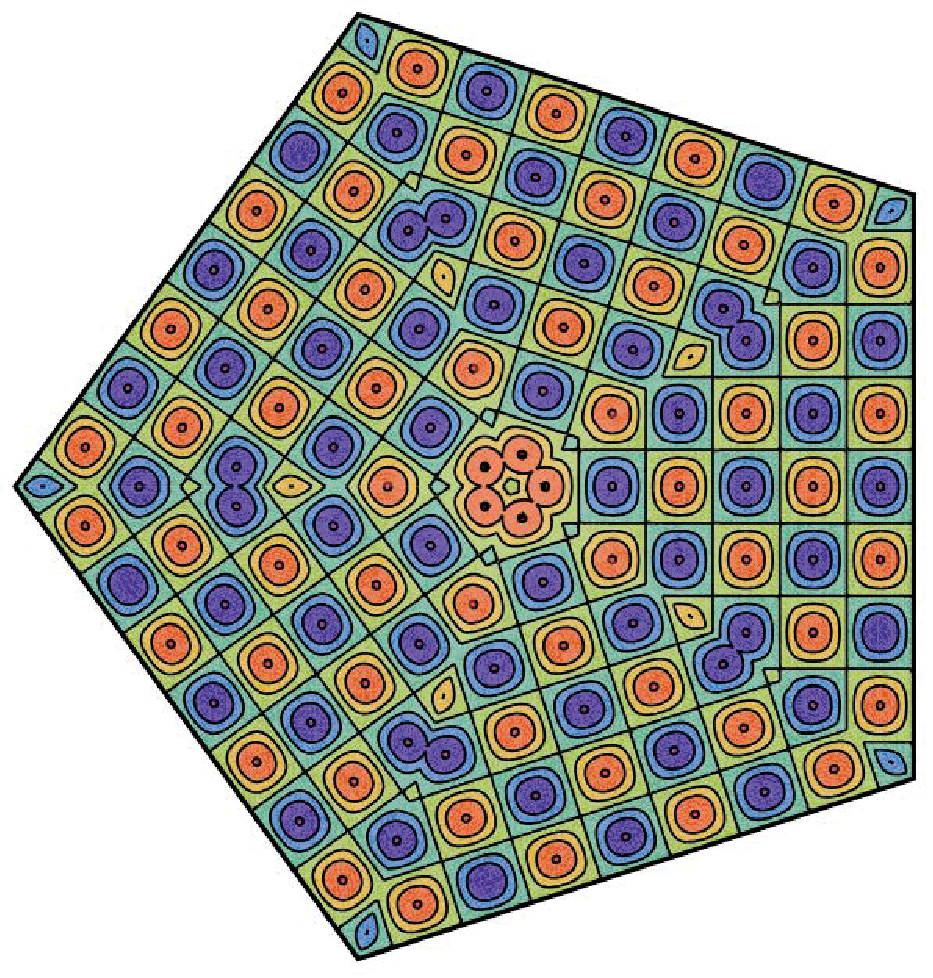} 
\caption{Plot of the normalized wave function from Eqs. (3) and (7) for the regular pentagonal quantum box with $m=3, n=2$. }
\label{fig10}}
\end{figure}
\begin{figure}
{\includegraphics[width=0.45\textwidth]{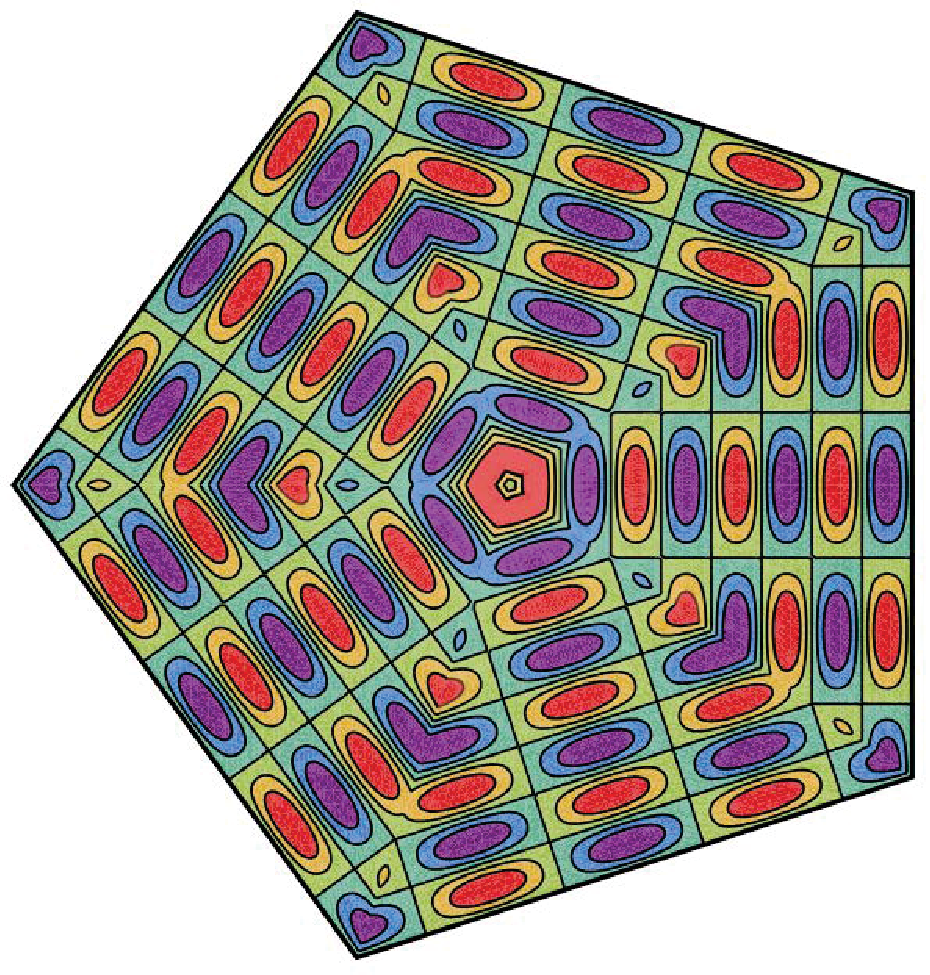} 
\caption{Plot of the normalized wave function from Eqs. (3) and (7) for the regular pentagonal quantum box with $m=4, n=2$. }
\label{fig11}}
\end{figure}
\begin{figure}
{\includegraphics[width=0.45\textwidth]{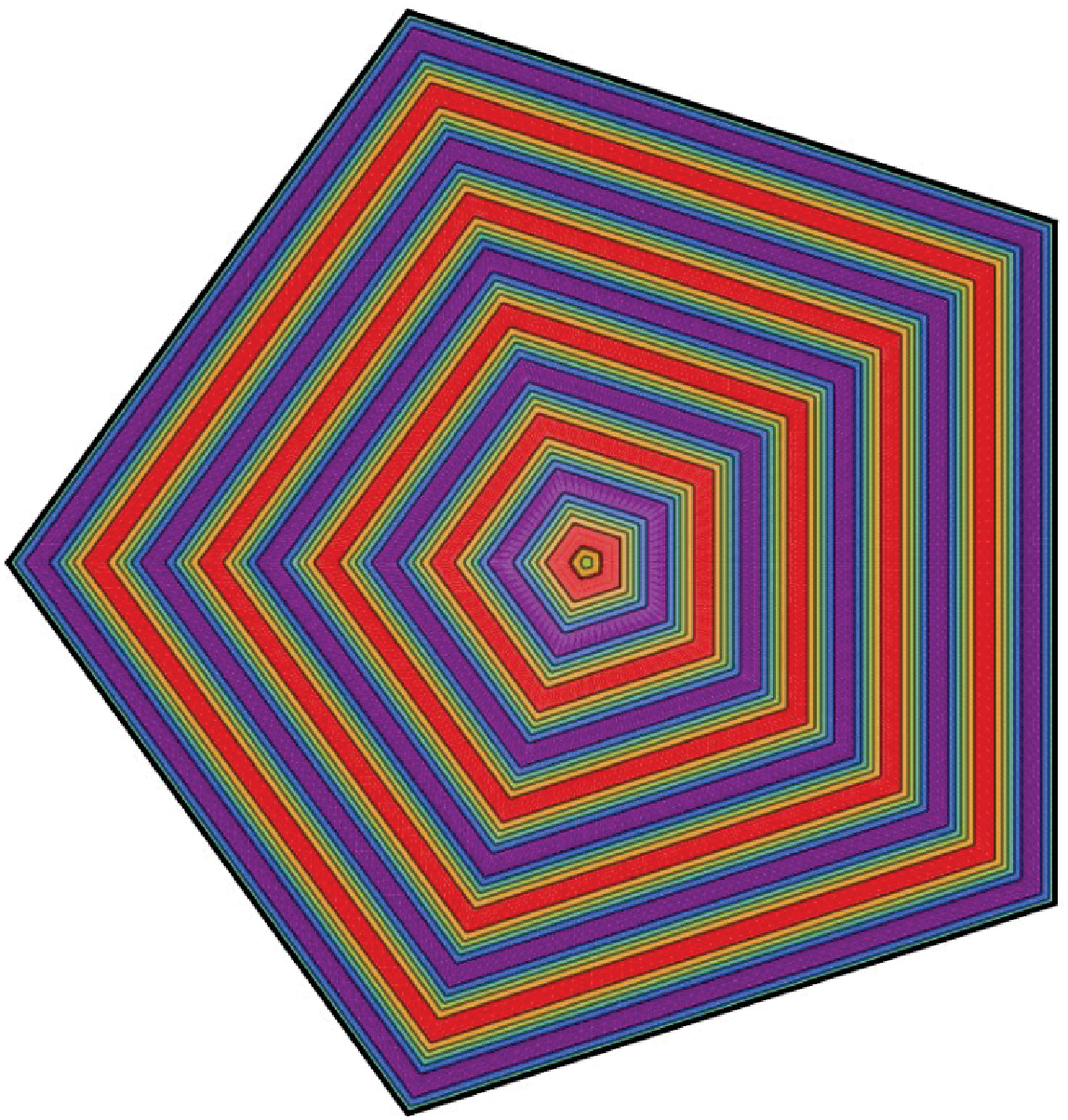} 
\caption{Plot of the normalized wave function from Eqs. (3) and (7) for the regular pentagonal quantum box with $m=5, n=2$. }
\label{fig12}}
\end{figure}
\vskip200pt
\section{Plots of the wave functions for $n=1,2,3$ of the thin  regular pentagonal microstrip antenna}
In order to protray the wave functions for the thin regular pentagonal microstrip antenna, the general form of which is given in Eq. (5), we used Mathematica to plot the isosceles triangular portion of pentagonal figure shown in Fig. 1, extending from the center at the origin to the corners at C and D, respectively.  then, using Mathematica, we added the green boundary along the vertical C-D line, and rotated each isosceles figure 4 times about the origin, making sure that the overall wave function was continuous at each point.  In order to do so, the odd wave function formula in Eq. (6) could not be used without cutting a slit from the origin to one of the corners. These wave functions will be the subject of another publication on the slitted regular pentagonal antenna.

Color plots of the wave functions for the regular pentagonal microstrip antenna with $n=1$ and $m=0$ to 5  are shown in Figs. (13)-(18).  Color plots of the wave functions for the regular pentagonal quatum box  with $n=2$ and $m=0$ to 5 in are shown in Figs. (19)-(24).  In each of these figures, the wave functions are continuous everywhere inside the box and due to the Dirichlet boundary condition that the wave functions all vanish on the boundaries, the pentagonal boundaries are portrayed in black.  Note that each of these 10 figures has the property that there are $nm$ concentric black regular pentagons, at which the wave functions vanish.  It is also noteworthy that since the boundary contains discontinuous derivatives at the five corner points, the wave functions also exhibit some additional discontinuous derivatives along the lines from the center to the five corners of the pentagonal boxes.
\begin{figure}
{\includegraphics[width=0.45\textwidth]{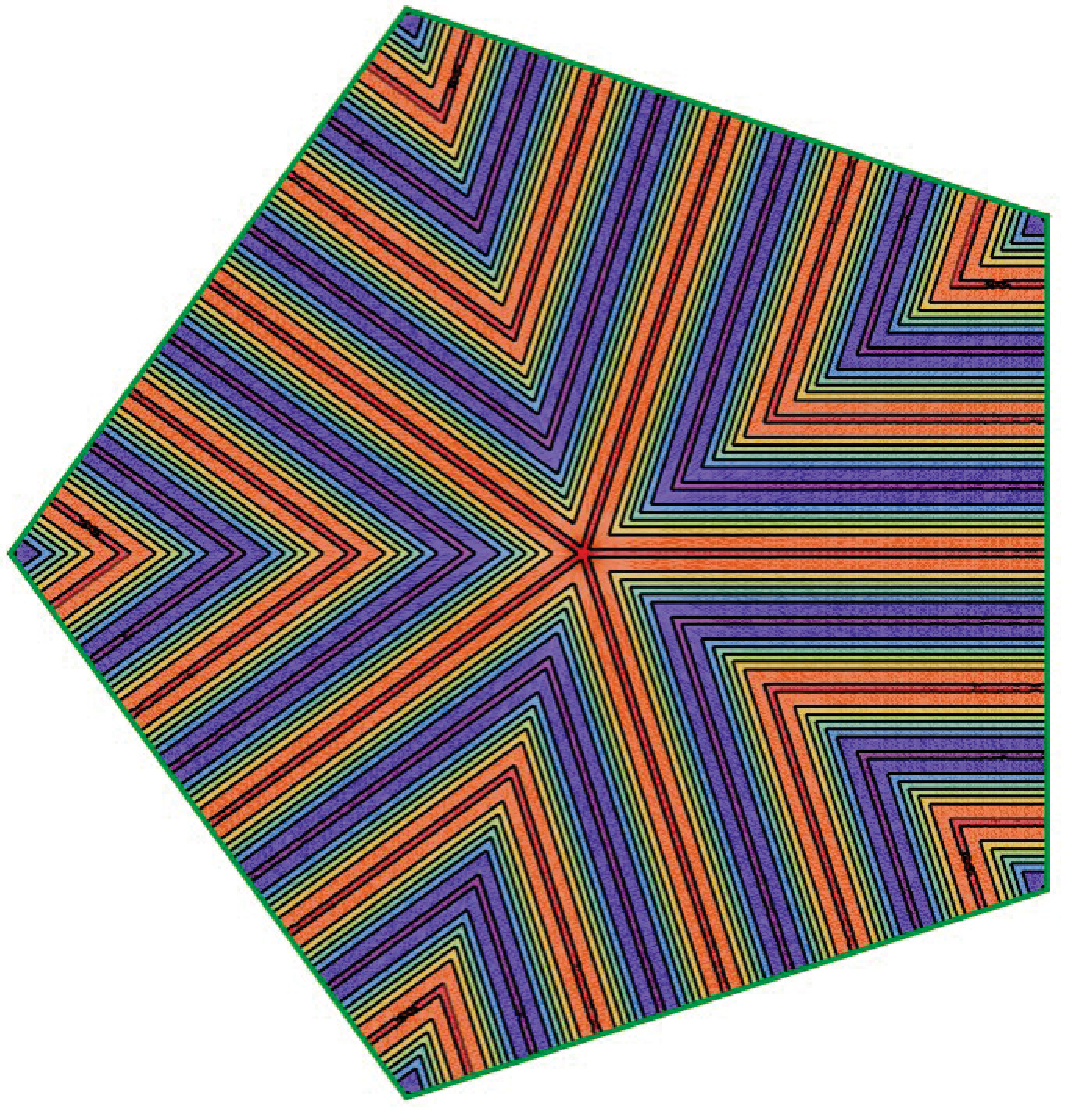} 
\caption{Plot of the normalized wave function from Eqs. (5) and (7) for the regular pentagonal microstrip antenna with $m=0, n=1$. }
\label{fig13}}
\end{figure} 
\begin{figure}
{\includegraphics[width=0.45\textwidth]{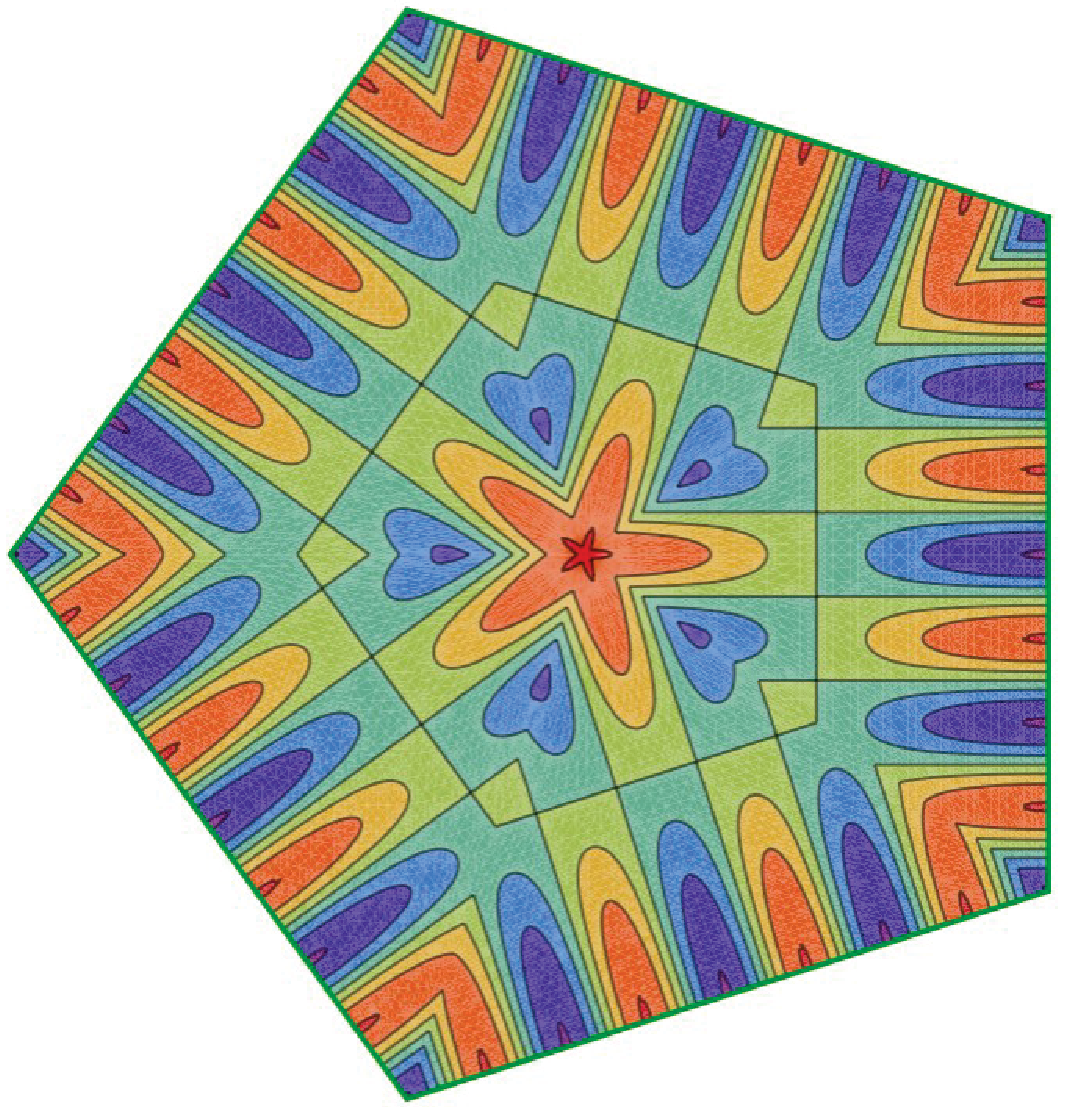} 
\caption{Plot of the normalized wave function from Eqs. (5) and (7) for the regular pentagonal microstrip antenna with $m=1,n=1$. }
\label{fig14}}
\end{figure}
\begin{figure}
{\includegraphics[width=0.45\textwidth]{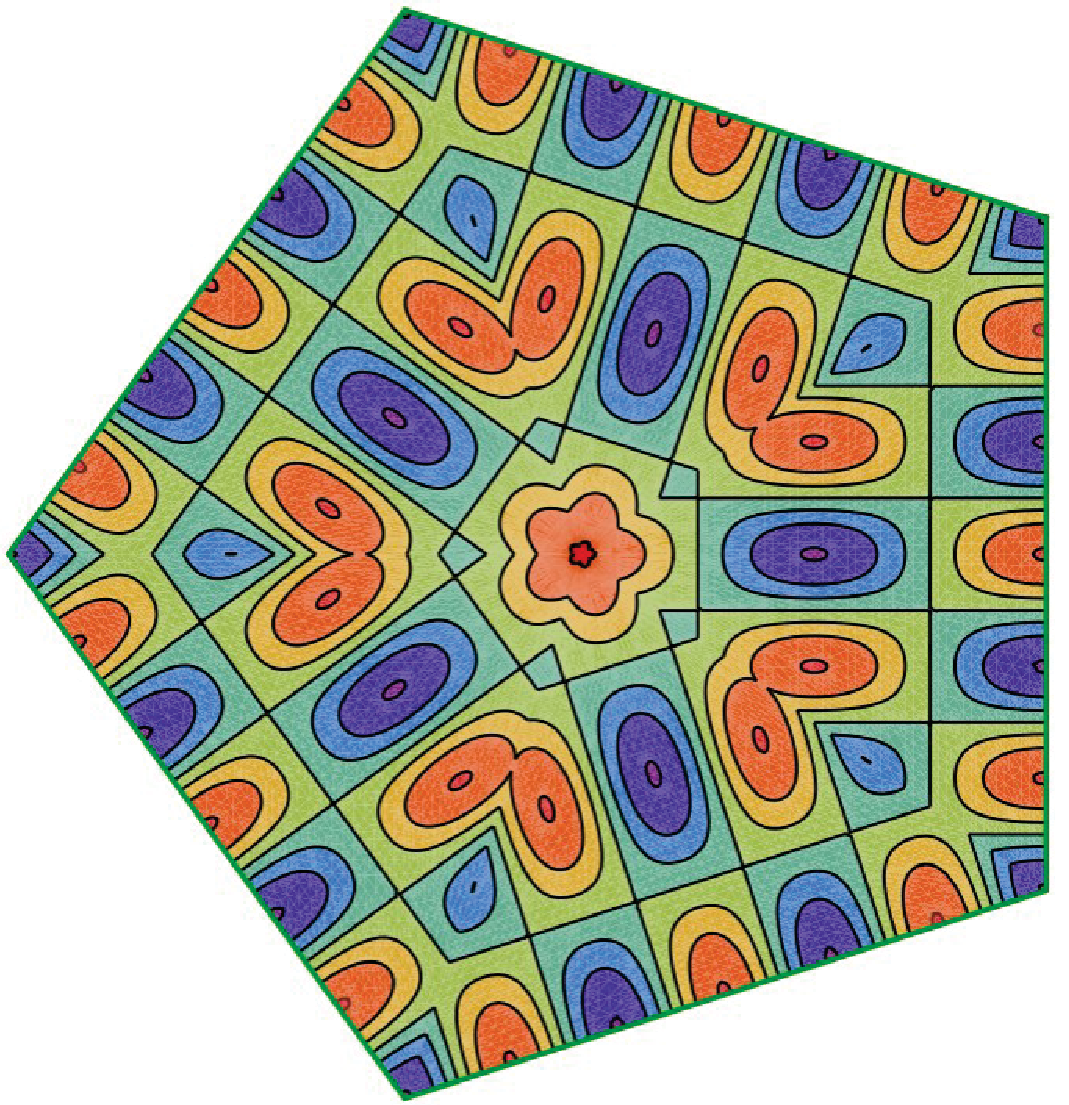} 
\caption{Plot of the normalized wave function from Eqs. (5) and (7) for the regular pentagonal microstrip antenna with $m=2, n=1$. }
\label{fig15}}
\end{figure} 
\begin{figure}
{\includegraphics[width=0.45\textwidth]{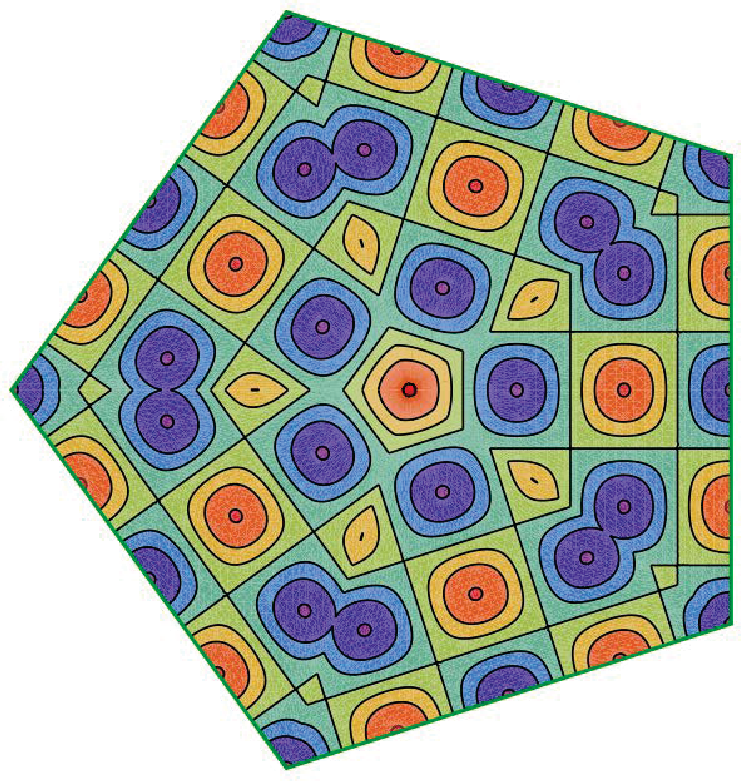} 
\caption{Plot of the normalized wave function from Eqs. (5) and (7) for the regular pentagonal microstrip antenna with $m=3, n=1$. }
\label{fig16}}
\end{figure} 
\begin{figure}
{\includegraphics[width=0.45\textwidth]{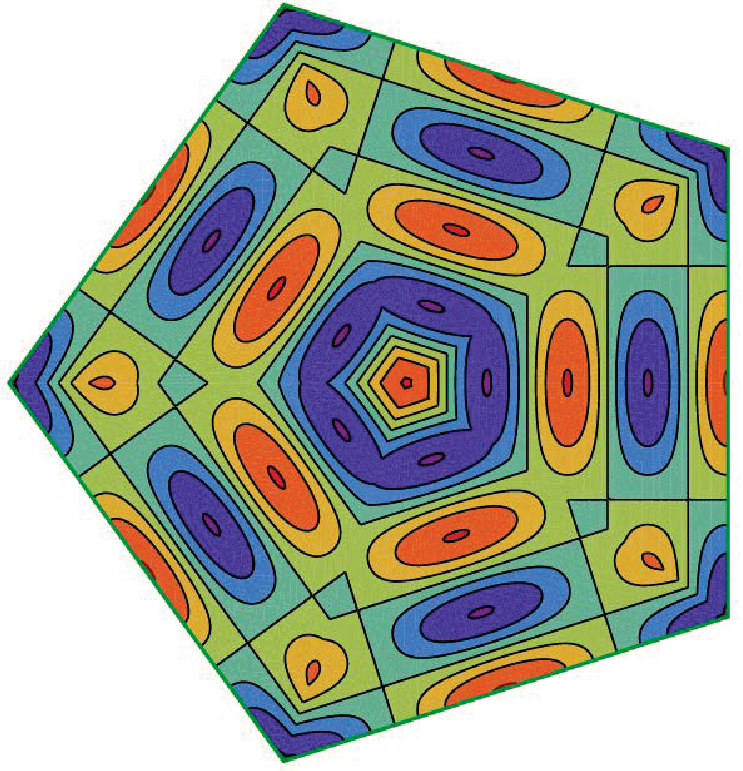} 
\caption{Plot of the normalized wave function from Eqs. (5) and (7) for the regular pentagonal  microstrip antenna with $m=4, n=1$. }
\label{fig17}}
\end{figure}
\begin{figure}
{\includegraphics[width=0.45\textwidth]{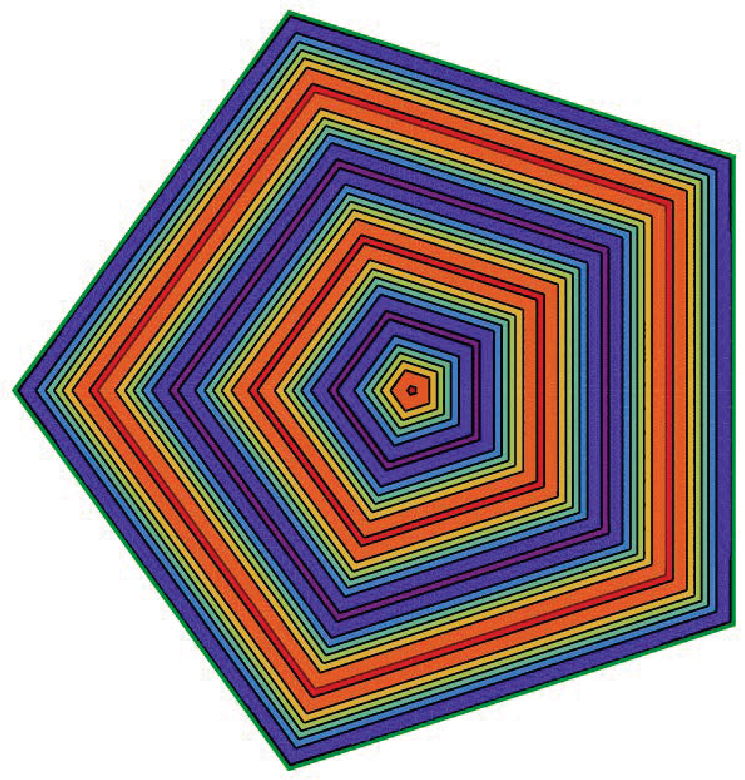} 
\caption{Plot of the normalized wave function from Eqs. (5) and (7) for the regular pentagonal  microstrip antenna with $m=5, n=1$. }
\label{fig18}}
\end{figure}
\begin{figure}
{\includegraphics[width=0.45\textwidth]{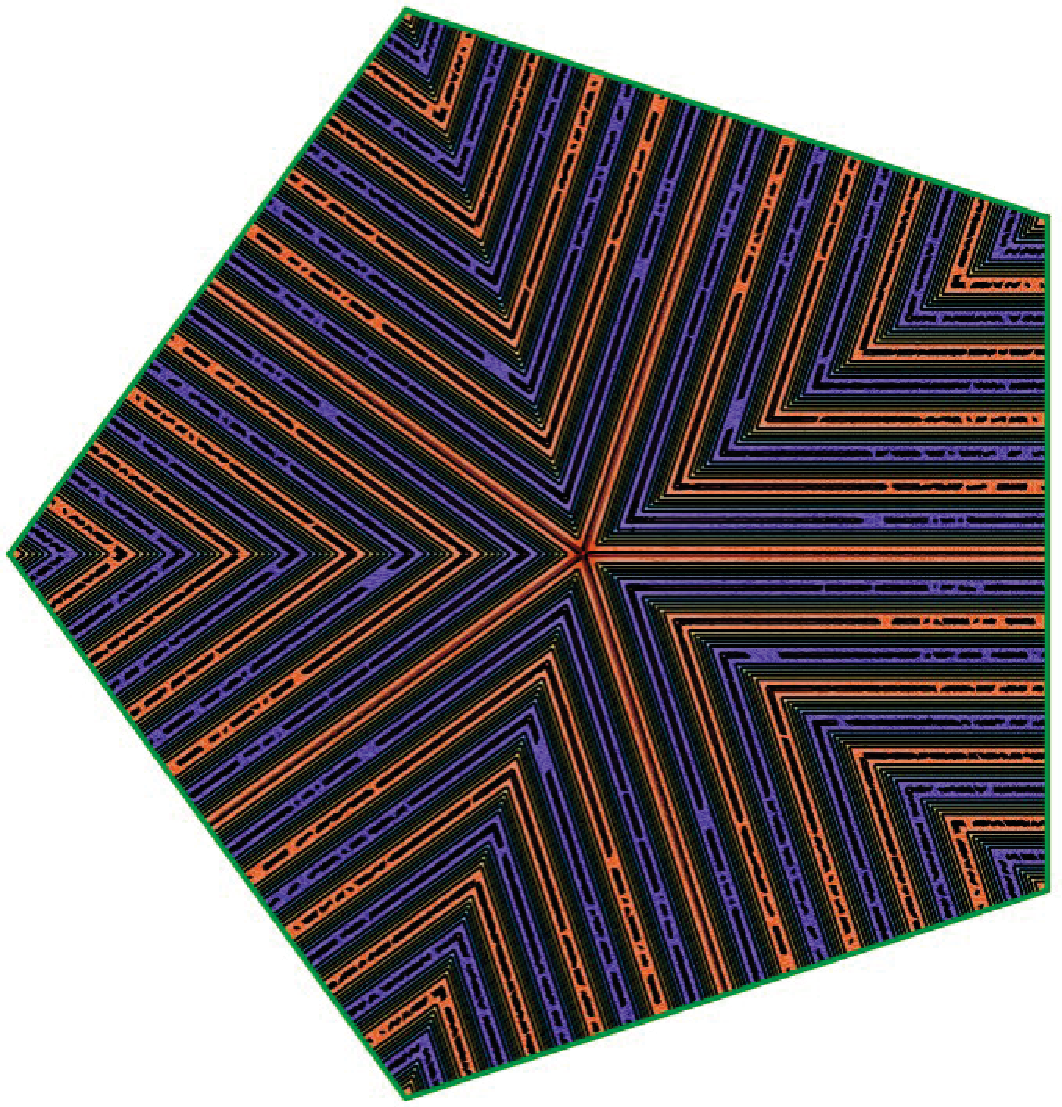} 
\caption{Plot of the normalized wave function from Eqs. (5) and (7) for the regular pentagonal microstrip antenna with $m=0, n=2$. }
\label{fig19}}
\end{figure} 
\begin{figure}
{\includegraphics[width=0.45\textwidth]{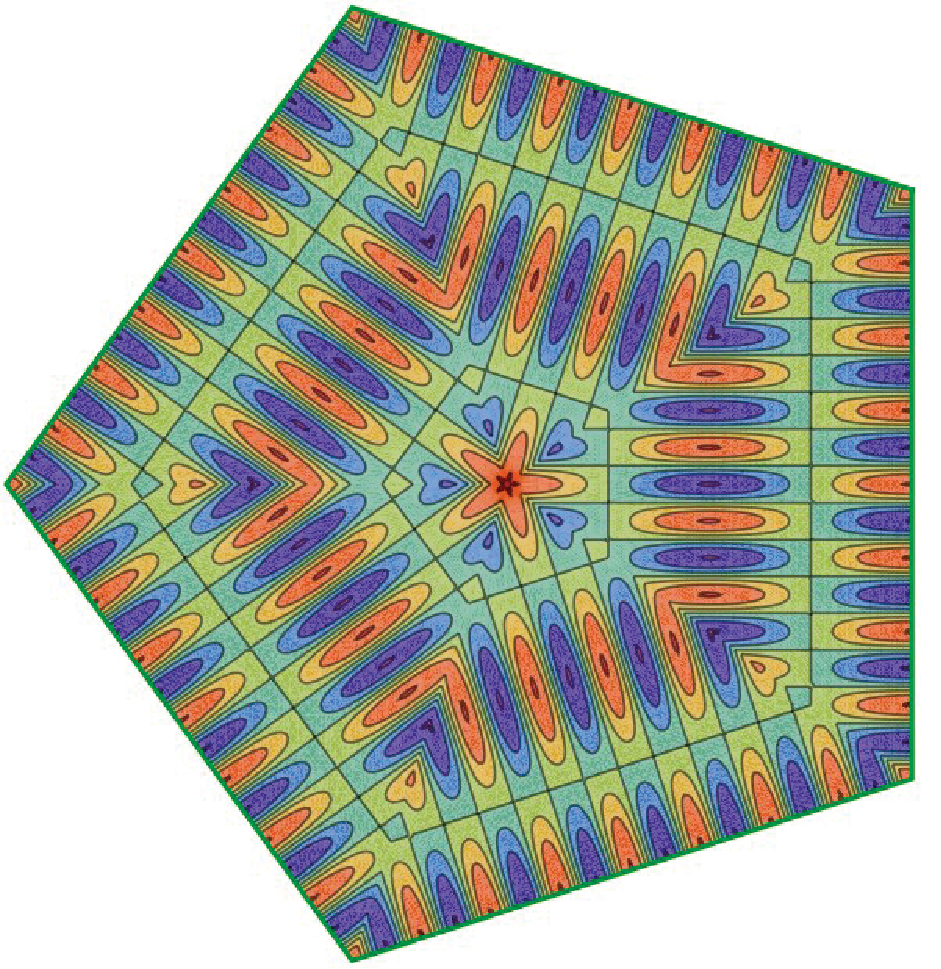} 
\caption{Plot of the normalized wave function from Eqs. (5) and (7) for the regular pentagonal microstrip antenna with $m=1, n=2$. }
\label{fig20}}
\end{figure} 
\begin{figure}
{\includegraphics[width=0.45\textwidth]{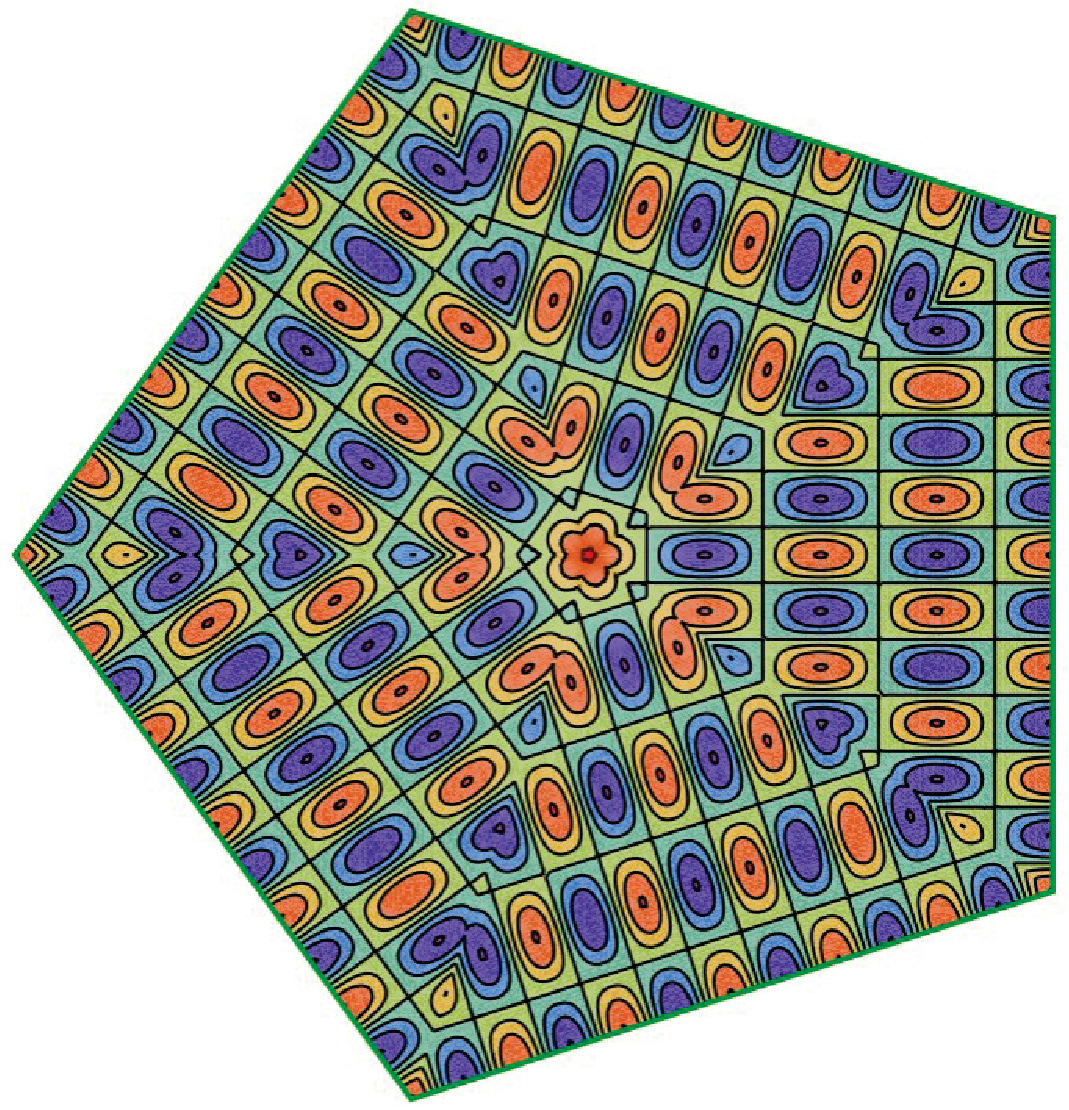} 
\caption{Plot of the normalized wave function from Eqs. (5) and (7) for the regular pentagonal microstrip antenna with $m=2, n=2 $. }
\label{fig21}}
\end{figure} 
\begin{figure}
{\includegraphics[width=0.45\textwidth]{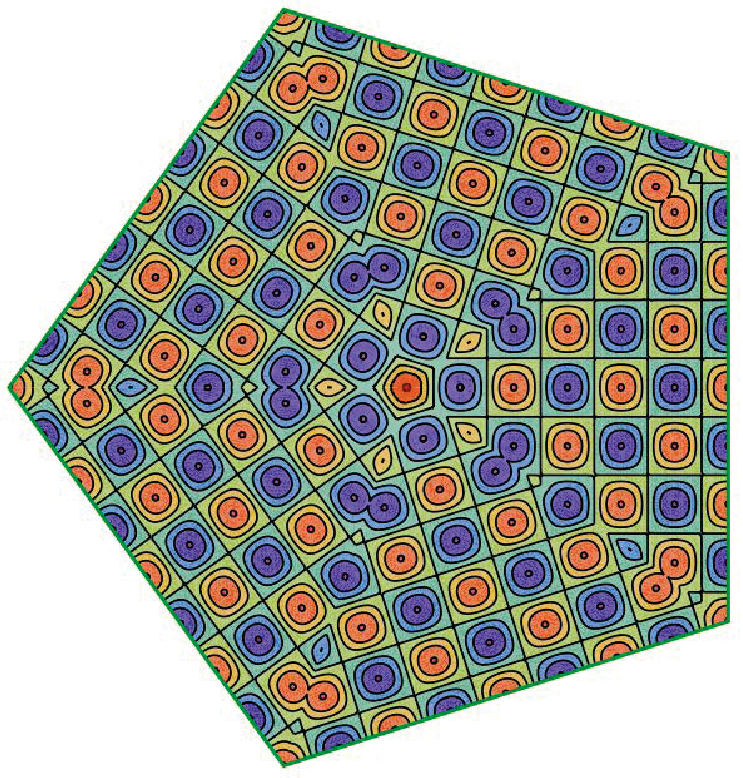} 
\caption{Plot of the normalized wavefunction from Eqs. (5) and (7) for the regular pentagonal microstrip antenna with $m=3, n=2$. }
\label{fig22}}
\end{figure} 
\begin{figure}
{\includegraphics[width=0.45\textwidth]{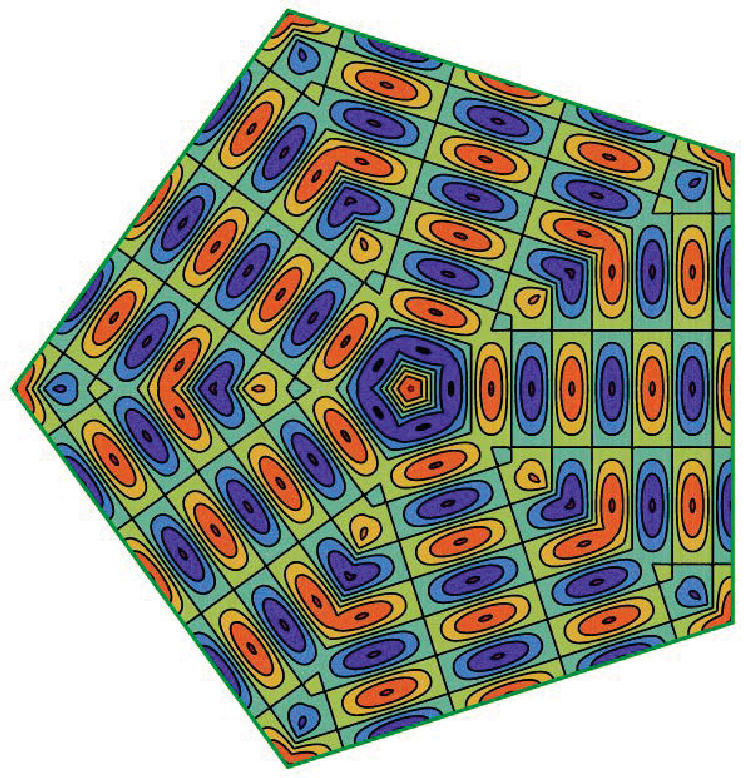} 
\caption{Plot of the normalized wave function from Eqs. (5) and (7) for the regular pentagonal microstrip antenna with $m=4, n=2$. }
\label{fig23}}
\end{figure}\begin{figure}
{\includegraphics[width=0.45\textwidth]{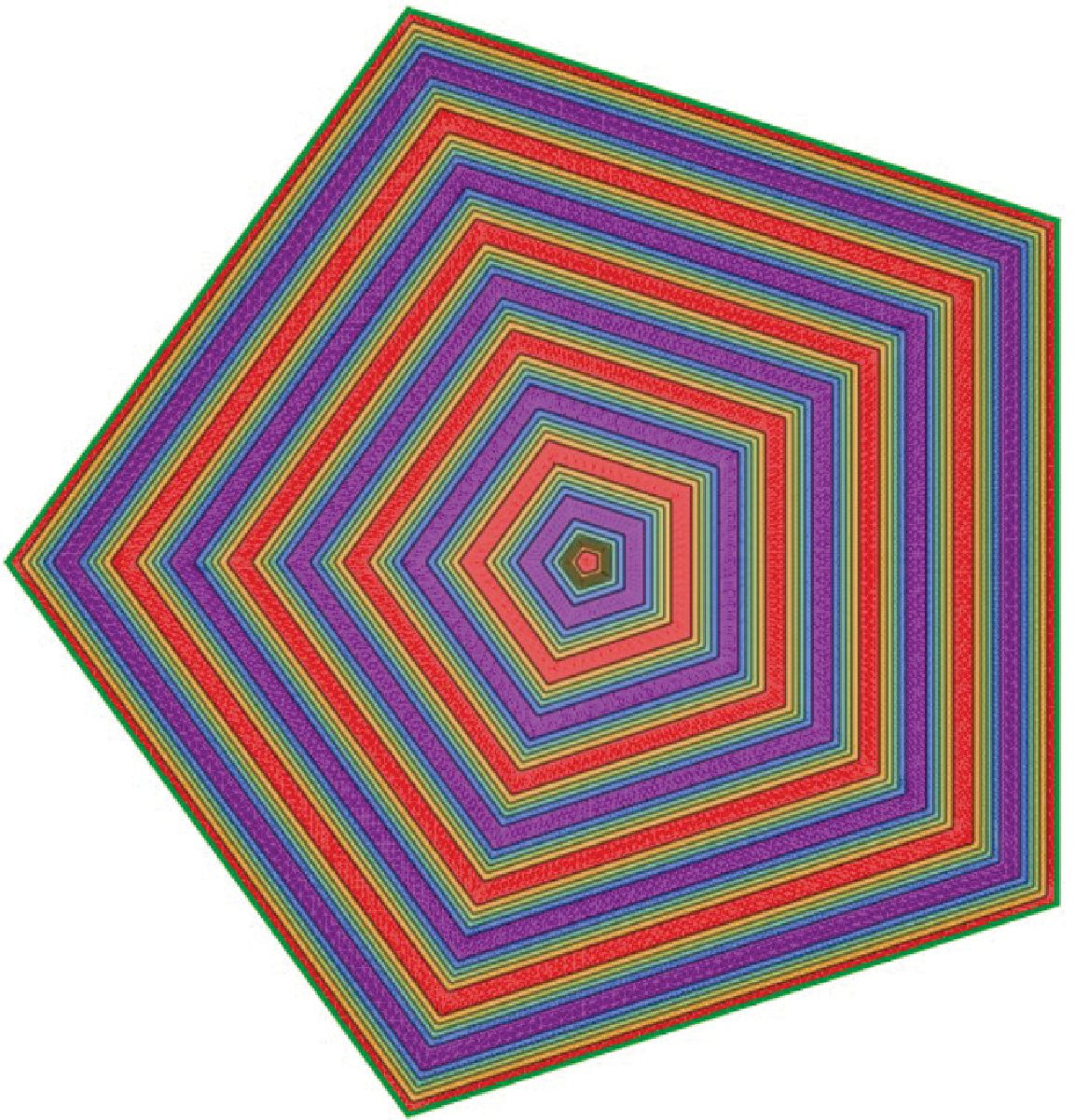} 
\caption{Plot of the normalized wave function from Eqs. (5) and (7) for the regular pentagonal microstrip antenna with $m=5, n=2$. }
\label{fig24}}
\end{figure}

\vskip30pt
\section{Appendix 1: Derivation of the general regular pentagonal box wave function}

For the regular pentagonal box, the wave function for the isosceles triangular region with corners at the origin and at points $C$ and $D$ that is an even function about the horizontal axis may be written as
\begin{eqnarray}
\frac{\Psi^{(e)}(x,y)}{N}&=&\sin\Bigl[\frac{\pi x\ell}{\alpha\cos(\pi/5)}\Bigr]\cos\Bigl[\frac{\pi y(m+n+o+p)}{\alpha\sin(\pi/5)}\Bigr]\nonumber\\
& &+\sin\Bigl[\frac{\pi x m}{\alpha\cos(\pi/5)}\Bigr]\cos\Bigl[\frac{\pi y(n+o+p+ \ell)}{\alpha\sin(\pi/5)}\Bigr]\nonumber\\
& &+\sin\Bigl[\frac{\pi x n}{\alpha\cos(\pi/5)}\Bigr]\cos\Bigl[\frac{\pi y(o+p+\ell+m)}{\alpha\sin(\pi/5)}\Bigr]\nonumber\\
& &+\sin\Bigl[\frac{\pi x o}{\alpha\cos(\pi/5)}\Bigr]\cos\Bigl[\frac{\pi y(p+\ell+m+n)}{\alpha\sin(\pi/5)}\Bigr]\nonumber\\
& &+\sin\Bigl[\frac{\pi x p}{\alpha\cos(\pi/5)}\Bigr]\cos\Bigl[\frac{\pi y(\ell+m+n+o)}{\alpha\sin(\pi/5)}\Bigr].\hskip18pt
\end{eqnarray}   
We note that when $x=\alpha\cos(\pi/5)$, each term in the expression for $\Psi^{(e)}$ vanishes, as required for this portion ($\frac{1}{5}$) of the quantum box.
                                    
Each of these five terms in Eq. (13) must satisfy the same Schr{\"o}dinger wave equation, Eq. (2).  Therefore after multiplying each term in the resulting Schrödinger equation for the wave function given in Eq. (13) by $\sin^2(\pi/5)\alpha^2/\pi^2$, we have 
\begin{eqnarray}
\ell^2\gamma+(m+n+o+p)^2  &=&m^2\gamma+(n+\ell+o+p)^2,\label{eqn3)}\\
&=&n^2\gamma+(\ell+m+o+p)^2,\hskip5pt\\
&=&o^2\gamma+(\ell+m+n+p)^2,\hskip5pt\\
&=&p^2\gamma+(\ell+m+n+o)^2,\hskip5pt
\end{eqnarray}  
where 
\begin{eqnarray}
\gamma&=&\tan^2(\pi/5)\\
&=&5-2\sqrt{5}.\label{eqn18}
\end{eqnarray}
After simplification of Eqs. (14)-(17), we may write
\begin{eqnarray}
 (m-\ell)[(m+\ell)(1-\gamma)+2(n+o+p)]&=&0,\\
 (n-\ell)[(n+\ell)(1-\gamma)+2(m+o+p)]&=&0,\\
 (o-\ell)[(o+\ell)(1-\gamma)+2(n+m+p)]&=&0,\\
 (p-\ell)[(p+\ell)(1-\gamma)+2(n+m+o)]&=&0.
 \end{eqnarray} 
 However, since from Eq. (18), $\gamma$ is an irrational number, and hence to satisfy Eqs. (14)-(17),  we require that  $\ell=m=n=o=p$.  Hence, the energy of this formulation of the  isosceles triangle that corresponds to 1/5 of the regular pentagon is only given by one quantum number.
 
 From Eq. (2), the energy $E$ of the wave function may be written as
\begin{eqnarray} 
 E&=&\frac{\hbar^2k^2}{2M},
 \end{eqnarray}
 where $k^2=k_x^2+k_y^2$.
 Then, from Eq. (13), we have
 \begin{eqnarray}
 \frac{\alpha^2k^2}{\pi^2}&=&\Bigl(\frac{\ell}{\cos(\pi/5)}\Bigr)^2+\Bigl(\frac{m+n+o+p}{\sin(\pi/5)}\Bigr)^2,\nonumber\\
 \end{eqnarray}
 and setting $\ell=m=o=p=n$, we have
 \begin{eqnarray}
 \frac{\alpha^2k^2_{n,1,4}}{\pi^2}&=&\frac{n^2}{\cos^2(\pi/5)}+\frac{(4n)^2}{\sin^2(\pi/5)},
  \end{eqnarray}
  and Eq. (13) may be rewritten as 
  \begin{eqnarray}
  \frac{\Psi^{(e)}_{1,n}(x,y)}{N}&=&\sin\Bigl[\frac{n\pi x}{\alpha\cos(\pi/5)}\Bigr]\cos\Bigl[\frac{4n\pi y}{\alpha\sin(\pi/5)}\Bigr].\nonumber\\
 \end{eqnarray} 
  Now, instead of Eq. (13), we could just as well have written 
\begin{eqnarray}
\frac{\Psi^{(e)}(x,y)}{N}&=&\sin\Bigl[\frac{\pi x(\ell+m)}{\alpha\cos(\pi/5)}\Bigr]\cos\Bigl[\frac{\pi y(n+o+p)}{\alpha\sin(\pi/5)}\Bigr]\nonumber\\
& &+\sin\Bigl[\frac{\pi x(m+n)}{\alpha\cos(\pi/5)}\Bigr]\cos\Bigl[\frac{\pi y(o+p+ \ell)}{\alpha\sin(\pi/5)}\Bigr]\nonumber\\
& &+\sin\Bigl[\frac{\pi x (n+o)}{\alpha\cos(\pi/5)}\Bigr]\cos\Bigl[\frac{\pi y(p+\ell+m)}{\alpha\sin(\pi/5)}\Bigr]\nonumber\\
& &+\sin\Bigl[\frac{\pi x(o+p)}{\alpha\cos(\pi/5)}\Bigr]\cos\Bigl[\frac{\pi y(\ell+m+n)}{\alpha\sin(\pi/5)}\Bigr]\nonumber\\
& &+\sin\Bigl[\frac{\pi x (p+\ell)}{\alpha\cos(\pi/5)}\Bigr]\cos\Bigl[\frac{\pi y(m+n+o)}{\alpha\sin(\pi/5)}\Bigr].\hskip20pt
\end{eqnarray}   
We note that when $x=\alpha\cos(\pi/5)$, each term in the expression for $\Psi^{(e)}$ vanishes, as required for this portion of the quantum box.
                                    
Since each of these five terms must satisfy the same Schr{\"o}dinger wave equation, Eq. (2).  
After multiplying each term  in the resulting Schr{\"o}dinger wave equation by $\sin^2(\pi/5)\alpha^2/\pi^2$, we have
\begin{eqnarray}
(\ell+m)^2\gamma+(n+o+p)^2  &=&(m+n)^2\gamma+(\ell+o+p)^2,\hskip20pt\\
&=&(n+o)^2\gamma+(\ell+m+p)^2,\hskip20pt\\
&=&(o+p)^2\gamma+(\ell+m+n)^2,\hskip20pt\\
&=&(p+\ell)^2\gamma+(m+n+o)^2.\hskip20pt
\end{eqnarray}  
From Eq. (29)-(32),  after simplification we may write
\begin{eqnarray}
(\ell-n)[\gamma(\ell+n+2m)-(n+\ell+2o+2p)]&=&0,\\
(n-p)[\gamma(n+p+2o)-(n+p+2\ell+2m)]&=&0,\\
 (m-p)[\gamma(m+p+2\ell)-(p+m+2n+2\ell)]&=&0,\\
 (o-\ell)[\gamma(o+\ell+2p)-(o+\ell+2m+2n)]&=&0.
 \end{eqnarray}
 These equations imply that $\ell=m=o=p=n$.
 But in this case, the energy of the isosceles triangle that corresponds to 1/5 of the regular pentagon is different than in the first case studied above.
 Instead of the wavevector in Eq. (26) and  the wavefunction in Eq. (27), we have for $\ell=m=n=o=p$,

   \begin{eqnarray}
\frac{\alpha^2k^2_{n,2,3}}{\pi^2}&=&\Bigl(\frac{2n}{\cos(\pi/5)}\Bigr)^2+\Bigl(\frac{3n}{\sin(\pi/5)}\Bigr)^2,\nonumber\\
  \frac{\Psi^{(e)}_{2,n}(x,y)}{N}&=&\sin\Bigl[\frac{2n\pi x}{\alpha\cos(\pi/5)}\Bigr]\cos\Bigl[\frac{3n\pi y}{\alpha\sin(\pi/5)}\Bigr].\nonumber\\
 \end{eqnarray} 

 This implies that there must be a second quantum number $m$. In order to prove the exact forms for the wavefunctions, we shall continue with the quantum regular pentagonal box wavefunction in a third form, differing from the forms presented in Eqs. (13) and (28),
 \begin{eqnarray}
\frac{\Psi^{(e)}(x,y)}{N}&=&\sin\Bigl[\frac{\pi x(\ell+m+n)}{\alpha\cos(\pi/5)}\Bigr]\cos\Bigl[\frac{\pi y(o+p)}{\alpha\sin(\pi/5)}\Bigr]\nonumber\\
& &+\sin\Bigl[\frac{\pi x(m+n+o)}{\alpha\cos(\pi/5)}\Bigr]\cos\Bigl[\frac{\pi y(p+ \ell)}{\alpha\sin(\pi/5)}\Bigr]\nonumber\\
& &+\sin\Bigl[\frac{\pi x (n+o+p)}{\alpha\cos(\pi/5)}\Bigr]\cos\Bigl[\frac{\pi y(\ell+m)}{\alpha\sin(\pi/5)}\Bigr]\nonumber\\
& &+\sin\Bigl[\frac{\pi x(o+p+\ell)}{\alpha\cos(\pi/5)}\Bigr]\cos\Bigl[\frac{\pi y(m+n)}{\alpha\sin(\pi/5)}\Bigr]\nonumber\\
& &+\sin\Bigl[\frac{\pi x (p+\ell+m)}{\alpha\cos(\pi/5)}\Bigr]\cos\Bigl[\frac{\pi y(n+o)}{\alpha\sin(\pi/5)}\Bigr].\hskip20pt
\end{eqnarray}   
  After multiplying each term in the Schrödinger wave equation by $\sin^2(\pi/5)\alpha^2/\pi^2$, we have
  \begin{eqnarray}
(\ell+m+n)^2\gamma+(o+p)^2  &=&(m+n+o)^2\gamma+(p+\ell)^2,\hskip20pt\\
&=&(n+o+p)^2\gamma+(\ell+m)^2,\hskip20pt\\
&=&(o+p+\ell)^2\gamma+(m+n)^2,\hskip20pt\\
&=&(p+\ell+m)^2\gamma+(n+o)^2.\hskip20pt
\end{eqnarray}   
Rewriting Eq. (37), the right sides of Eqs. (37) and (38), the right sides of Eqs. (38) and (39), and the left side of Eq. (37) in combination with Eq. (40), we have
\begin{eqnarray}
(\ell-o)\left\{[\ell+o+2(m+n)]\gamma-(o+\ell+2p)\right\}&=&0,\\
(m-p)[(n+o)\gamma-\ell]&=&0,\\
(n-\ell)[(n+\ell+2o+2p)\gamma-\ell-n-2m]&=&0,\\
(n-p)[(n+p+2\ell+2m)\gamma-p-n-2o]&=&0.
\end{eqnarray}
These equations are consistent with $\ell=m=o=p=n$, so that

 \begin{eqnarray}
\frac{\alpha^2k^2_{n,3,2}}{\pi^2}&=&\Bigl(\frac{3n}{\cos(\pi/5)}\Bigr)^2+\Bigl(\frac{2n}{\sin(\pi/5)}\Bigr)^2,\nonumber\\
 \frac{\Psi^{(e)}_{3,n}(x,y)}{N}&=&\sin\Bigl[\frac{3n\pi x}{\alpha\cos(\pi/5)}\Bigr]\cos\Bigl[\frac{2n\pi y}{\alpha\sin(\pi/5)}\Bigr].\nonumber\\
 \end{eqnarray}
 Then, we consider a possible fourth form for the box wave function,
  \begin{eqnarray}
\frac{\Psi^{(e)}(x,y)}{N}&=&\sin\Bigl[\frac{\pi x(\ell+m+n+o)}{\alpha\cos(\pi/5)}\Bigr]\cos\Bigl[\frac{\pi yp}{\alpha\sin(\pi/5)}\Bigr]\nonumber\\
& &+\sin\Bigl[\frac{\pi x(m+n+o+p)}{\alpha\cos(\pi/5)}\Bigr]\cos\Bigl[\frac{\pi y\ell}{\alpha\sin(\pi/5)}\Bigr]\nonumber\\
& &+\sin\Bigl[\frac{\pi x (n+o+p+\ell)}{\alpha\cos(\pi/5)}\Bigr]\cos\Bigl[\frac{\pi ym}{\alpha\sin(\pi/5)}\Bigr]\nonumber\\
& &+\sin\Bigl[\frac{\pi x(o+p+\ell+m)}{\alpha\cos(\pi/5)}\Bigr]\cos\Bigl[\frac{\pi yn}{\alpha\sin(\pi/5)}\Bigr]\nonumber\\
& &+\sin\Bigl[\frac{\pi x (p+\ell+m+n)}{\alpha\cos(\pi/5)}\Bigr]\cos\Bigl[\frac{\pi yo}{\alpha\sin(\pi/5)}\Bigr].\hskip20pt
\end{eqnarray} 
As above,  after multiplying each term in the Schrödinger wave equation by $\sin^2(\pi/5)\alpha^2/\pi^2$, we have
  \begin{eqnarray}
(\ell+m+n+o)^2\gamma+p^2  &=&(m+n+o+p)^2\gamma+\ell^2,\hskip20pt\\
&=&(n+o+p+\ell)^2\gamma+m^2,\hskip20pt\\
&=&(o+p+\ell+m)^2\gamma+n^2,\hskip20pt\\
&=&(p+\ell+m+n)^2\gamma+o^2.\hskip20pt
\end{eqnarray}   
From Eqs. (49)- (52), we have
\begin{eqnarray}
(\ell-p)[(\ell+p)(1+\gamma)+2(m+n+o)\gamma]&=&0,\hskip20pt\\
(\ell-m)[(\ell+m)(1-\gamma)-2\gamma(n+o+p)]&=&0,\hskip20pt\\
(n-m)[(n+m)(1-\gamma)-2\gamma(o+p+\ell)]&=&0,\hskip20pt\\
(n-o)[(n+o)(1-\gamma)-2\gamma(p+\ell+m)]&=&0.\hskip20pt
\end{eqnarray}
These equations are all consistent with $\ell=m=n=o=p$, so that 
 \begin{eqnarray}
\frac{\alpha^2k^2_{n,4,1}}{\pi^2}&=&\Bigl(\frac{4n}{\cos(\pi/5)}\Bigr)^2+\Bigl(\frac{n}{\sin(\pi/5)}\Bigr)^2,\nonumber\\
\frac{\Psi^{(e)}_{4,n}(x,y)}{N}&=&\sin\Bigl[\frac{4n\pi x}{\alpha\cos(\pi/5)}\Bigr]\cos\Bigl[\frac{n\pi y}{\alpha\sin(\pi/5)}\Bigr].\nonumber\\
 \end{eqnarray}
Finally, the fifth possible forms for the  wavevector and  wave function for the box with  $\ell=m=n=o=p$ is 
 \begin{eqnarray}
\frac{\alpha^2k^2_{n,5,0}}{\pi^2}&=&\Bigl(\frac{5n}{\cos(\pi/5)}\Bigr)^2,\nonumber\\
\frac{\Psi^{(e)}_{5,n}(x,y)}{N}&=&\sin\Bigl[\frac{5n\pi x}{\alpha\cos(\pi/5)}\Bigr].\nonumber\\
 \end{eqnarray}
 
 Therefore, the precise form of the general regular pentagonal box wave function given in Eq. (3) for the isosceles triangular region depicted in Fig. 1 has been derived for general $(n,m)$, where $n\ge1$ is unbounded and $0\le m\le5$.  It is straightforward to do the same for the regular pentagonal microstrip antenna wave function presented in Eq. (4). 
 \section{Appendix 2:  Normalizing the general regular pentagonal quantum box wave function}

Now, we need to check the normalization for the general $\Psi^{(e,o)}_{n,m}(x,y)$ wave functions in the horizontal isosceles triangle, for both the antenna and the box.  For the even box, we have
\begin{eqnarray}
\frac{1}{5}&=&\biggl(A^{(e)}_{nm}\biggr)^2\int_0^{\alpha\cos(\pi/5)}\sin^2\Biggl(\frac{nm\pi x}{\alpha\cos(\pi/5)}\Biggr)dx\times\nonumber\\
& &\int_{-x\tan(\pi/5)}^{x\tan(\pi/5)}dy\cos^2\Biggl(\frac{(5-m)n\pi y}{\alpha\sin(\pi/5)}\Biggr).
\end{eqnarray}
Letting
\begin{eqnarray}
x'&=&\frac{nm\pi x}{\alpha\cos(\pi/5)},\\
y'&=&\frac{(5-m)n\pi y}{\alpha\sin(\pi/5)},
\end{eqnarray}
and we obtain
\begin{eqnarray}
\frac{1}{5}&=&2\biggl(A^{(e)}_{nm}\biggr)^2\frac{\alpha^2\sin(\pi/5)\cos(\pi/5)}{n^2\pi^2m(5-m)}\int_0^{nm\pi}\sin^2(x')dx'\nonumber\\
& &\times\int_{0}^{(5-m)x'/m}dy'\cos^2(y').\\
&=&2\biggl(A^{(e)}_{n,m}\biggr)^2\frac{\alpha^2\sin(\pi/5)\cos(\pi/5)}{n^2\pi^2m(5-m)}\int_0^{nm\pi}\sin^2(x')dx'\nonumber\\
& &\times \int_{0}^{(5-m)x'/m}dy'\frac{1}{2}[1+\cos(2y')].\\
&=&2\biggl(A^{(e)}_{n,m}\biggr)^2\frac{\alpha^2\sin(\pi/5)\cos(\pi/5)}{n^2\pi^2m(5-m)}\int_0^{nm\pi}\sin^2(x')dx'\nonumber\\
& &\times\Biggl[\frac{x'}{2}\Bigl(\frac{5-m}{m}\Bigr)+\frac{1}{4}\sin\Bigl(\frac{2(5-m)x'}{m}\Bigr)\Biggr].\\
&=&\biggl(A^{(e)}_{n,m}\biggr)^2\frac{\alpha^2\sin(\pi/5)\cos(\pi/5)}{n^2\pi^2m(5-m)}\int_0^{nm\pi}[1-\cos(2x')]dx'\nonumber\\
& &\times\Biggl[\frac{x'}{2}\Bigl(\frac{5-m}{m}\Bigr)+\frac{1}{4}\sin\Bigl(\frac{2(5-m)x'}{m}\Bigr)\Biggr]\\
&=&2\biggl(A^{(e)}_{n,m}\biggr)^2\frac{\alpha^2\sin(\pi/5)\cos(\pi/5)}{n^2\pi^2m(5-m)}\frac{(5-m)n^2\pi^2m}{8}\\
&=&\frac{\alpha^2\sin(\pi/5)\cos(\pi/5)}{4}\Bigl(A_{n,m}^{(e)}\Bigr)^2.
\end{eqnarray}
Therefore, we have
\begin{eqnarray}
A^{(e)}_{n,m}=N=\frac{4}{a}\sqrt{\frac{\tan(\pi/5)}{5}},
\end{eqnarray}
which is independent of $n$ and $m$, where we used Eq. (1).  Similarly, it can be shown that $A^{(o)}=A^{(e)}$ is also independent of $n$ and $m$.  This is true for both the pentagonal box and the pentagonal antenna in Eqs. (3)-(6).

\section{Appendix 3: Mathematica\copyright  code for the $n=1, m=1$ antenna wave function plot}
\vskip5pt
borderThickness = 0.5\\
borderShade = 0.6 (*0=black, 1=green*)\\
\vskip5pt
waveFunctionPlot =\\
  ContourPlot[(2/(Sqrt[5*Sin[Pi/5]*Cos[Pi/5]]))*\\
   Cos[Pi*x/Cos[Pi/5]]*Cos[4*Pi*y/Sin[Pi/5]],\\
   \{x,0,Cos[Pi/5]\},\{y,-x*Tan[Pi/5],x*Tan[Pi/5]\},\\
   ColorFunction $\rightarrow$ ''Rainbow'',Axes $\rightarrow$ False,\\
   Frame $\rightarrow$ False, PlotRange $\rightarrow$ All,\\
   PerformanceGoal $\rightarrow$ ''Quality'']; (*n=1,m=1*)\\
   \\
   baseGraphics = First@Normal[waveFunctionPlot];\\
   \\
   outerVertex = \{Cos[Pi/5],Cos[Pi/5]*Tan[Pi/5]\};\\
   pentagonVertices =\\
    Table[RotationTrasnform[n*2*Pi/5,\{0,0\}][outerVertex],\\
    \{n,0,4\}];\\
    \\
    Graphics[\{Table[\\
     GeometricTransformation[baseGraphics,\\
     RotationTransform[n*2*Pi/5,\{0,0\}]],\{n,0,4\}],\{Thickness[\\
     borderThickness/100.0],\\
     Directive[Blend[\{Black,Green\},borderShade]],\\
     Line[Append[pentagonVertices,First[pentagonVertices]]]\}\},\\
    PlotRange $\rightarrow$ All, Axes $\rightarrow$ False, Frame $\rightarrow$ False]\\

\vskip30pt
\section{Appendix 4: Wave functions for the regular pentagonal antenna with $n=3$ and $m=0$ to $5$}
\begin{figure}
{\includegraphics[width=0.45\textwidth]{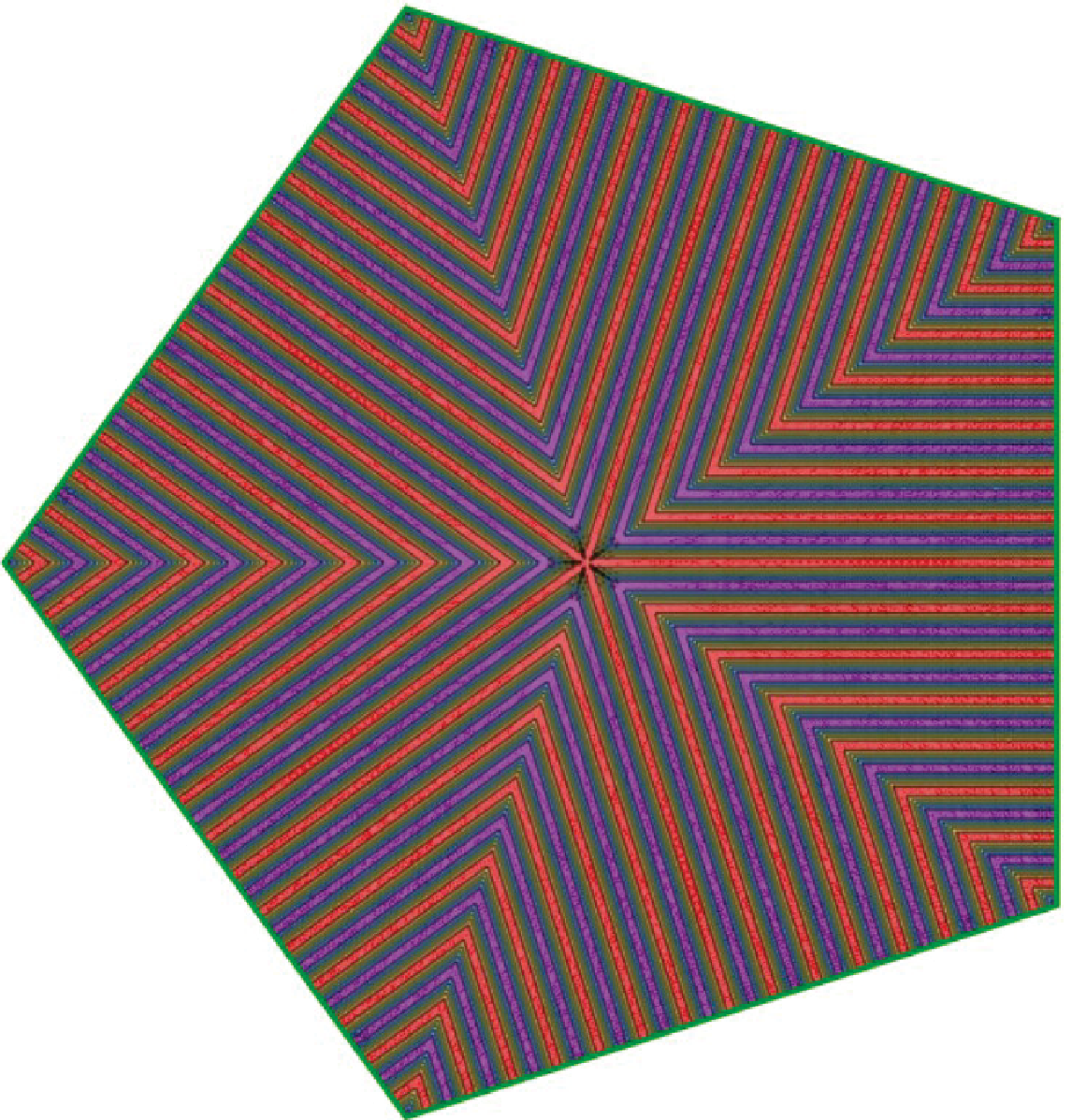} 
\caption{Plot of the normalized wave function from Eqs. (5) and (7) for the regular pentagonal microstrip antenna with $m=0, n=3$. }
\label{fig30}}
\end{figure} 
\begin{figure}
{\includegraphics[width=0.45\textwidth]{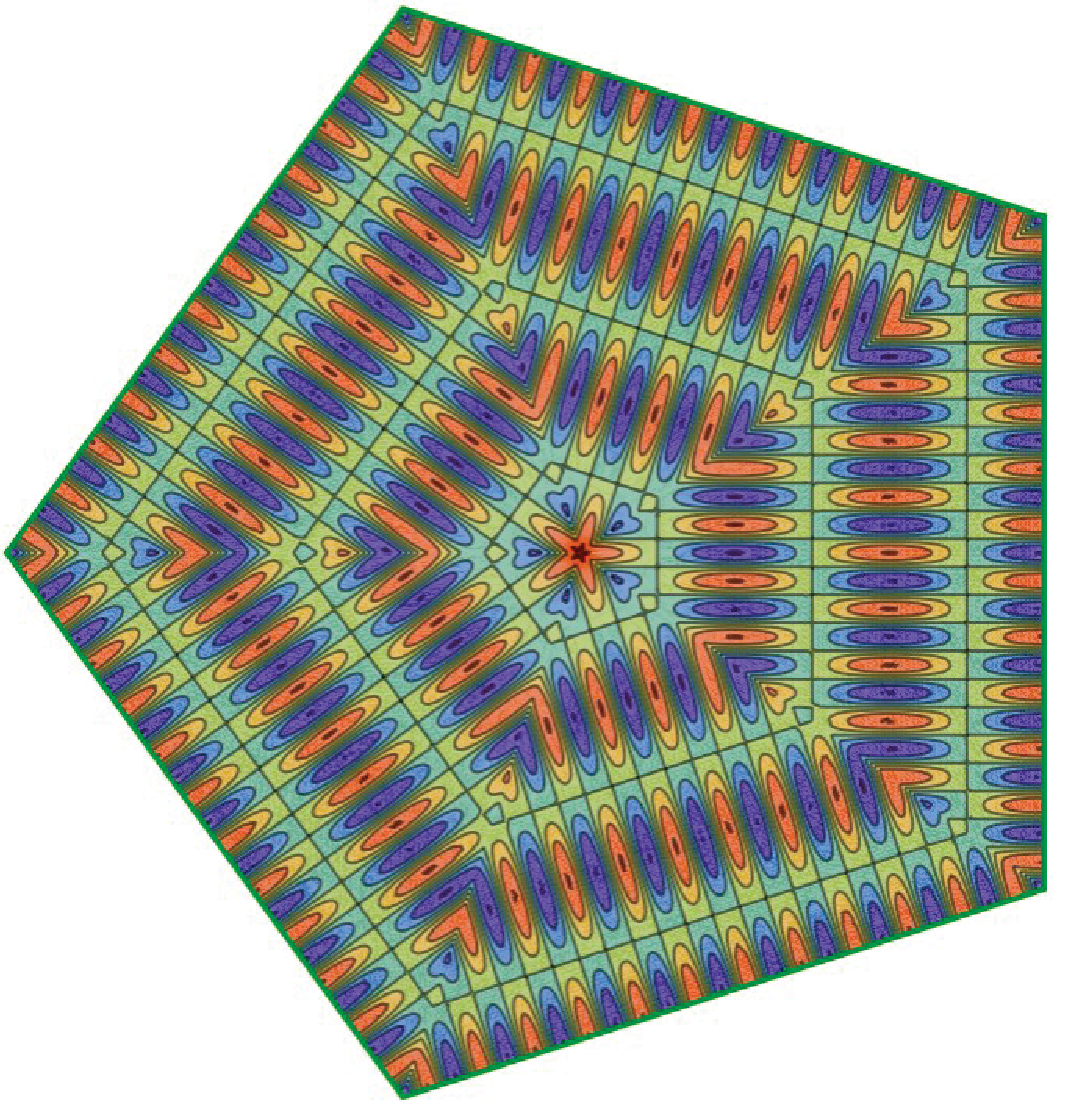} 
\caption{Plot of the normalized wave function from Eqs. (5) and (7) for the regular pentagonal microstrip antenna with $m=1, n=3$. }
\label{fig31}}
\end{figure}
\begin{figure}
{\includegraphics[width=0.45\textwidth]{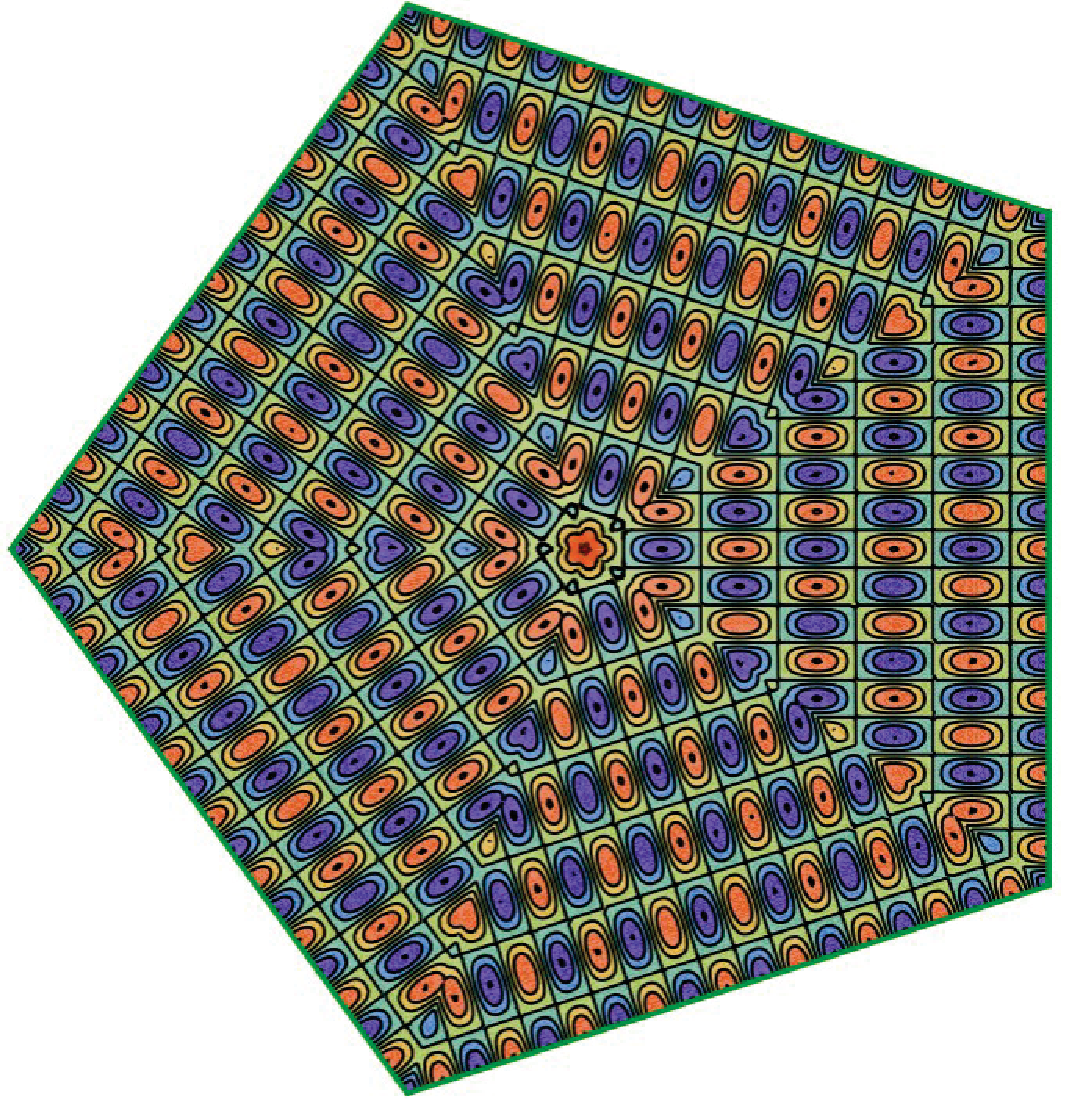} 
\caption{Plot of the normalized wave function from Eqs. (5) and (7) for the regular pentagonal microstrip antenna with $m=2, n=3$. }
\label{fig32}}
\end{figure} 
\begin{figure}
{\includegraphics[width=0.45\textwidth]{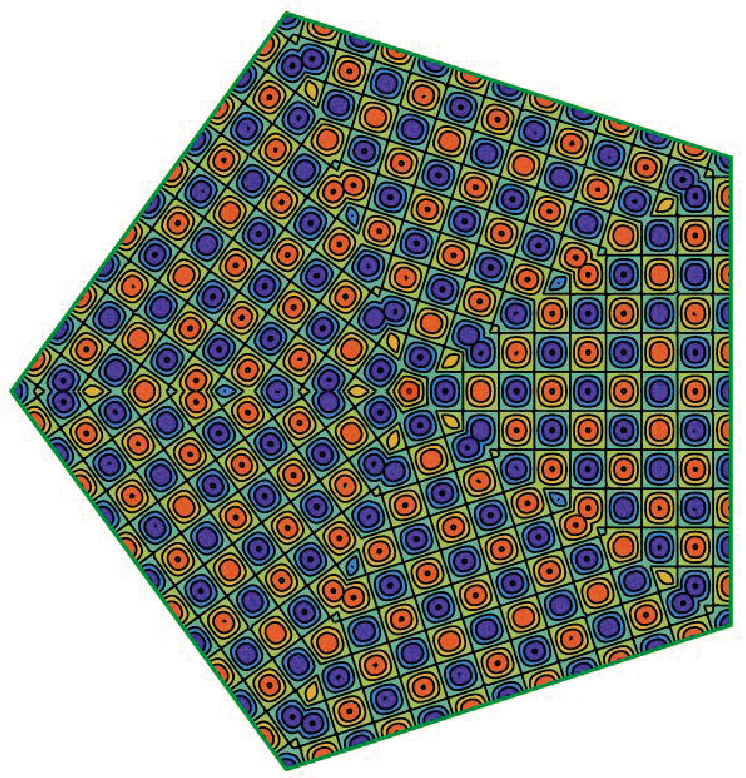} 
\caption{Plot of the normalized wave function from Eqs. (5) and (7) for the regular pentagonal microstrip antenna with $m=3, n=3$. }
\label{fig33}}
\end{figure} 
\begin{figure}
{\includegraphics[width=0.45\textwidth]{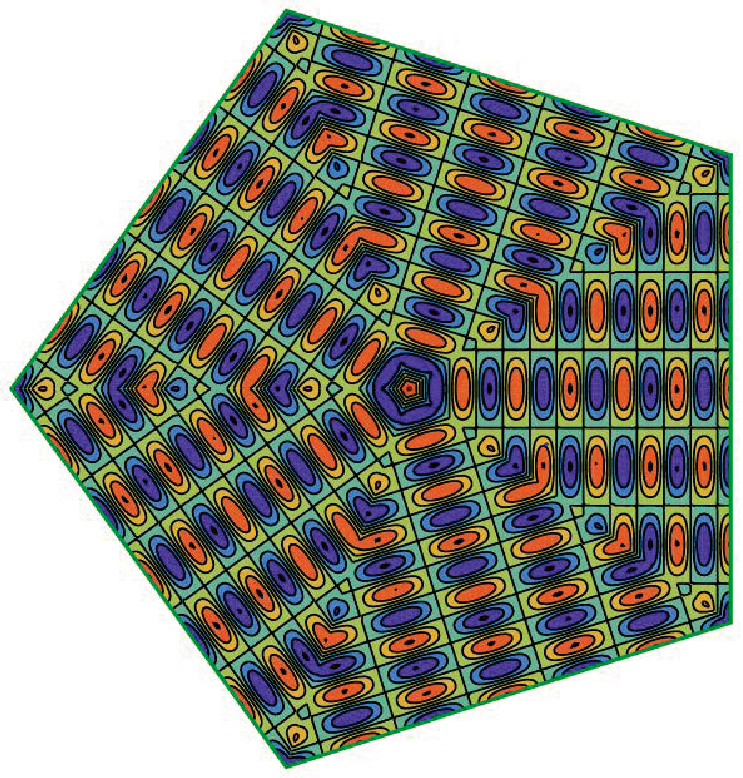} 
\caption{Plot of the normalized wave function from Eqs. (5) and (7) for the regular pentagonal  microstrip antenna with $m=4, n=3$. }
\label{fig34}}
\end{figure}
\begin{figure}
{\includegraphics[width=0.45\textwidth]{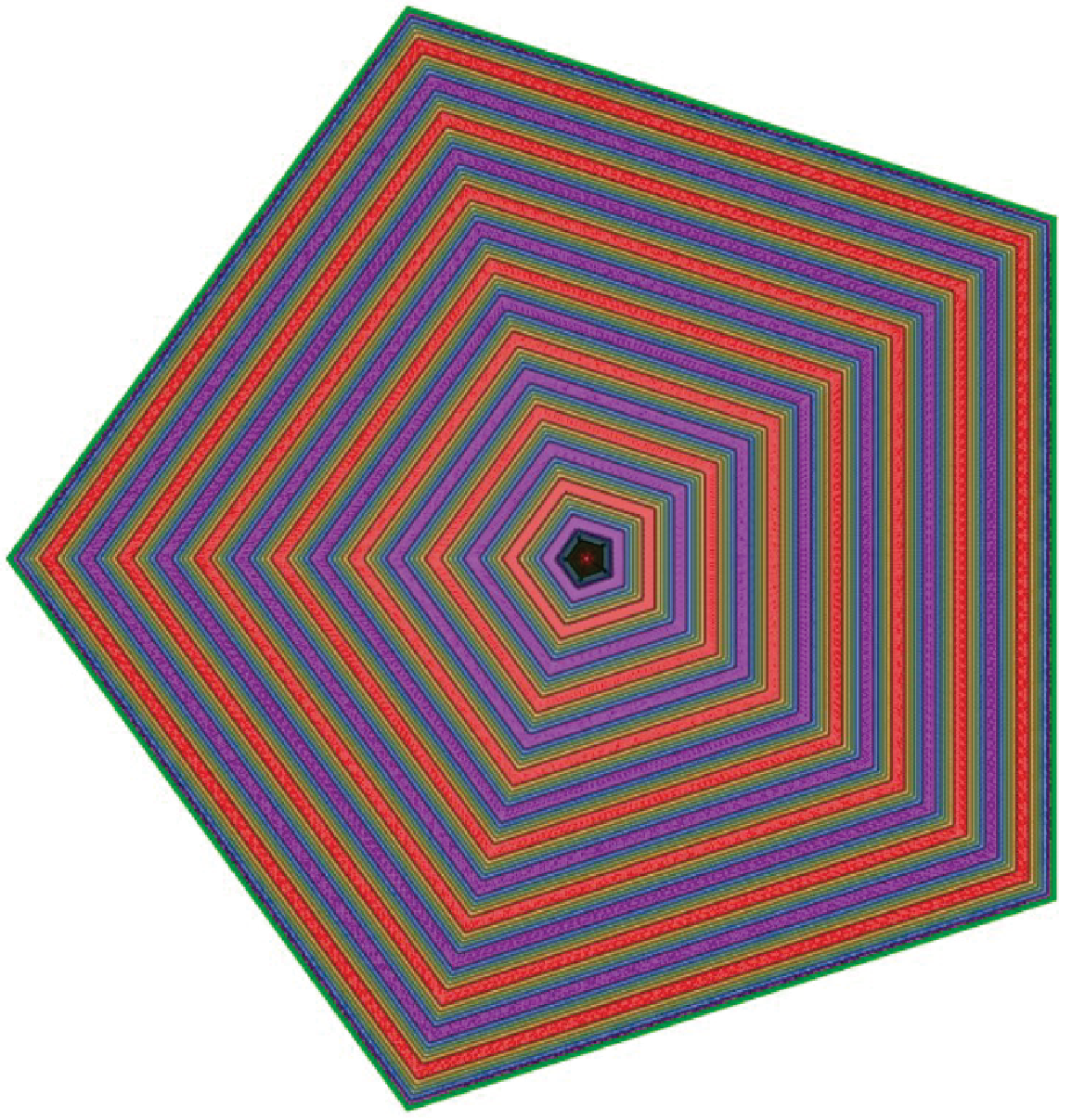} 
\caption{Plot of the normalized wave function from Eqs. (5) and (7) for the regular pentagonal  microstrip antenna with $m=5, n=3$. }
\label{fig35}}
\end{figure}

\vskip30pt

     \section{Summary and Conclusions}
    We found the exact forms of the wave functions for the two-dimensional quantum regular pentagonal quantum box, in which a single particle obeying the Schrödinger wave equation without a potential energy has a wave function that vanishes on the pentagonal boundary of the box, the Dirichlet condition, and to the thin microstrip regular pentagonal microstrip antenna, for which the electromagnetic field satisfies the Schrödinger wave equation without a potential energy, the normal derivative of which vanishes on the pentagonal boundary, the Neumann condition.  We found that there are two quantum numbers $n$ and $m$, where $n\ge1$ is unbounded, but $m_{\rm min}\le m\le 5$, where $m_{\rm min}=0$ for the antenna and $m_{\rm min}=1$ for the box. Although there are two exact forms appropriate for the wave functions corresponding to  each of the two boundary conditions, continuity of the wave functions along each of the five lines from the center to each of the corners of the pentagon reduces the number of exact forms for each boundary condition to just one.  Normalizing each form using the criterion that the probability of finding an effective single particle in the pentagon is one resulted in a normalization constant that is independent of both quantum numbers, as is the same value for the box and the antenna.

    \section{Acknowledgments}
The authors thank Kazuo Kadowaki and Kaveh Delfanazari for stimulating discussions.  T. L., E. E. D., and J. O. G. contributed equally to this work.  R. A. K. supervised the work and wrote the paper.

\section{References}
\noindent{[1]}. R. A. Klemm, E. R. LaBerge, D. R. Morley, T. Kashiwagi, M. Tsujimoto,  and K. Kadowaki, Cavity mode waves during terahertz radiation from rectangular Bi$_2$Sr$_2$CaCu$_2$O$_{8+\delta}$  mesas, J. Phys.: Condens. Matter {\bf 23}, 025701 (2011). \\
\noindent{[2]}. D. P. Cerkoney, C. Reid, C. M. Doty, A. Gramajo, T.  D. Campbell, M. A. Morales, K. Delfanazari, M. Tsujimoto, T. Kashiwagi, T. Yamamoto, C. Watanabe, H. Minami, K. Kadowaki, and R. A. Klemm,  Cavity mode enhancement of terahertz emission from equilateral triangular microstrip antennas of the high-$T_c$ superconductor Bi$_2$Sr$_2$CaCu$_2$O$_{8+\delta}$,   J. Phys.: Condens. Matter {\bf 29}, 015601 (2017). \\
\noindent{[3]}. R. A. Klemm, A. E. Davis, and Q. X. Wang,  Terahertz emission from thermally-managed square intrinsic Josephson junction microstrip antennas, IEEE J. of Sel. Top. Quant. Electron. {\bf 23} (4), Art. No. 8501208 (2017).\\
\noindent{[4]}. J. R. Rain,  P. Cai,  A. Baekey, M. A. Reinhard, R. I. Vasquez, A. C. Silverman,  C. L. Cain, and R. A. Klemm,   Wave functions for high-symmetry thin microstrip antennas and two-dimensional quantum boxes, Phys. Rev. A {\bf 104}, 062205 (2021).\\
\noindent{[5]}. K. Kadowaki, M. Tsujimoto, K. Yamaki, T. Yamamoto, T. Kashiwagi, H. Minami, M. Tachiki, and R. A. Klemm,  Evidence for a dual-source mechanism of THz radiation from rectangular mesas of single crystalline Bi$_2$Sr$_2$CaCu$_2$O$_{8+\delta}$  intrinsic Josephson junctions, J. Phys. Soc. Jpn. {\bf 79}, 023703 (2010).\\
\noindent{[6]}. R. A. Klemm and K. Kadowaki,  Angular dependence of the radiation power of a Josephson STAR-emitter,   J. Supercond.  Nov. Magn. {\bf 23}, 613 (2010).  \\
\noindent{[7]}.   M. Tsujimoto, K. Yamaki, K. Deguchi, T. Yamamoto, T. Kashiwagi, H. Minami, M. Tachiki, K. Kadowaki, and R. A. Klemm,  Geometrical resonance conditions for THz radiation from the intrinsic Josephson junctions in Bi$_2$Sr$_2$CaCu$_2$O$_{8+\delta}$, Phys. Rev. Lett. {\bf 105}, 037005 (2010). \\
\noindent{[8]}. R. A. Klemm and K. Kadowaki,  Output from a Josephson stimulated terahertz amplified radiation emitter,  J. Phys.: Condens. Matter {\bf 22}, 375701 (2010).\\
\noindent{[9]} M. Tinkham, {\it Group Theory and Quantum Mechanics} (McGraw-Hill, New York 1964).\\
\noindent{[10]}. T. Kashiwagi, K. Yamaki, M. Tsujimoto, K. Deguchi, N. Orita, T. Koike, R. Nakayama, H. Minami, T. Yamamoto, R. A. Klemm, M. Tachiki, and K. Kadowaki,  Geometrical full-wavelength resonance mode generating terahertz waves from a single crystalline Bi$_2$Sr$_2$CaCa$_2$O$_{8+\delta}$  rectangular mesa,  J. Phys. Soc. Jpn. {\bf 80}, 094709 (2011). \\
\noindent{[11]}. T. M. Benseman, A. E. Koshelev, K. E. Gray,  W.-K. Kwok, U. Welp, K. Kadowaki, M. Tachiki,  and T. Yamamoto, Tunable terahertz emission from Bi$_2$Sr$_2$CaCu$_2$O$_{8+\delta}$  mesa devices, Phys. Rev. B {\bf 84}, 064523 (2011).\\
\noindent{[12]}. T. Kashiwagi, M. Tsujimoto, T. Yamamoto, H. Minami, K. Yamaki, K. Delfanazari, K. Deguchi, N. Orita, T. Koike, R. Nakayama, T. Kitamura, M. Sawamura, S. Hagino, K. Ishida, K. Ivanovic, H. Asai, M. Tachiki, R. A. Klemm, and K. Kadowaki,  High temperature superconductor terahertz emitters:  Fundamental physics and its applications, Jap. J. Appl. Phys. {\bf 51}, 010113 (2012). \\
\noindent{[13]}. M. Tsujimoto, T. Yamamoto, K. Delfanazari, R. Nakayama, T. Kitamura, M. Sawamura,  T. Kashiwagi, H. Minami, M. Tachiki, K. Kadowaki, and R. A. Klemm,  Broadly tunable sub-terahertz emission from internal current-voltage characteristic branches generated from Bi$_2$Sr$_2$CaCu$_2$O$_{8+\delta}$  single crystals, Phys. Rev. Lett. {\bf 108}, 107006 (2012).\\
\noindent{[14]}. K. Delfanazari, H. Asai, M. Tsujimoto, T. Kashiwagi, T. Kitamura, T. Yamamoto, M. Sawamura, K. Ishida, R. A. Klemm, T. Hattori, and K. Kadowaki,  Tunable terahertz emission from triangular mesas of layered high-$T_c$ superconducting Bi$_2$Sr$_2$CaCu$_2$O$_{8+\delta}$  intrinsic Josephson junctions,  Opt. Express {\bf 21}, 2171 (2013).\\
\noindent{[15]}. T. M. Benseman, K. E. Gray, A. E. Koshelev, W.-K. Kwok, U. Welp, H. Mnami, K. Kadowaki, and T. Yamamoto,  Powerful terahertz emission from   Bi$_2$Sr$_2$CaCu$_2$O$_{8+\delta}$  mesa arrays, Appl. Phys. Lett. {\bf 103} 022602 (2013).\\
\noindent{[16]}. S. Sekimoto, C. Watanabe, H. Minami, T. Yamamoto, T. Kashiwagi, R. A. Klemm, and K. Kadowaki,  Continuous 30$\mu$W terahertz source by a high-$T_c$ superconductor mesa structure, Appl. Phys. Lett. {\bf 103}, 182601 (2013).\\
\noindent{[17]}.  K. Delfanazari, H. Asai, M. Tsujimoto, T. Kashiwagi, T. Kitamura,  K. Ishida, C. Watanabe, S. Sekimoto, T. Yamamoto, H. Minami, M. Tachiki, R. A. Klemm, T. Hattori, and K. Kadowaki, Terahertz oscillating devices based upon the intrinsic Josephson junctions in a high temperature superconductor, J. Infrared Milli Terahz Waves {\bf 35}, 131-146 (2014). \\
\noindent{[18]}. H. Minami, C. Watanabe, K. Sato, S. Sekimoto, T. Yamamoto, T. Kashiwagi, R. A. Klemm, and K. Kadowaki,  Local SiC photoluminescence evidence of  hot spot formation and sub-THz coherent emission from a rectangular Bi$_2$Sr$_2$CaCu$_2$O$_{8+\delta}$  mesa,  Phys. Rev. B {\bf 89}, 054503 (2014).\\
\noindent{[19]}. T. Kashiwagi, K. Nakade, Y. Saiwai, H. Minami, T. Kitamura, C. Watanabe, K. Ishida, S. Sekimoto,
K. Asanuma, T. Yasui, Y. Shibano, M. Tsujimoto, T. Yamamoto, B. Markovi{\'c}, J. Mirkovi{\'c}, R. A. Klemm and K. Kadowaki, 
Computed tomography image using sub-terahertz waves generated from a high-$T_c$ superconducting intrinsic Josephson junction oscillator,  Appl. Phys. Lett. {\bf 104}, 082603 (2014).\\
\noindent{[20]}. C. Watanabe, H. Minami, T. Yamamoto, T. Kashiwagi, R. A. Klemm, and K. Kadowaki, Spectral investigation of hot spot and cavity resonance effects on the terahertz radiation emitted from high-$T_c$ superconducting Bi$_2$Sr$_2$CaCu$_2$O$_{8+\delta}$  mesas, J. Phys. Condens. Matter {\bf 26}, 172201 (2014). \\
\noindent{[21]}. T. Kitamura, T. Kashiwagi, T. Yamamoto, C. Watanabe, K. Ishida, S. Sekimoto, K. Asanuma, T. Yasui, K. Nakade, Y. Shibano, Y. Saiwai, H. Minami, R. A. Klemm, and K. Kadowaki,  Broadly tunable, high-power terahertz radiation up to 73 K from a stand-alone Bi$_2$Sr$_2$CaCu$_2$O$_{8+\delta}$  mesa, Appl. Phys. Lett. {\bf 105}, 202603 (2014). \\
\noindent{[22]}. C. Watanabe, H. Minami, T. Kitamura, K. Asanuma, K. Nakade, T. Yasui, Y. Saiwai, Y. Shibano, T. Yamamoto, T. Kashiwagi, R. A. Klemm, and K. Kadowaki,  Influence of the local heating position on the terahertz emission power from high-T$_c$ superconducting Bi$_2$Sr$_2$CaCu$_2$O$_{8+\delta}$  mesas,  Appl. Phys. Lett. {\bf 106}, 042603 (2015).\\
\noindent{[23]}. T. Kashiwagi, T. Yamamoto, T. Kitamura, K. Asanuma, C. Watanabe, K. Nakade, T. Yasui, Y. Saiwai, Y. Shibano, H. Kubo, K. Sakamoto, T. Katsuragawa, M. Tsujimoto,  K. Delfanazari, R. Yoshizaki, H. Minami, R. A. Klemm, and K. Kadowaki,  Generation of electromagnetic waves from 0.3 to 1.6 THz  with a high-$T_c$ superconducting Bi$_2$Sr$_2$CaCu$_2$O$_{8+\delta}$   intrinsic Josephson junction emitter,  Appl. Phys. Lett. {\bf 106}, 092601 (2015).\\
\noindent{[24]}. K. Delfanazari, H. Asai, M. Tsujimoto, T. Kashiwagi, T. Kitamura, T. Yamamoto, W. Wilson, R. A. Klemm, T. Hattori, and K. Kadowaki,  Effect of bias electrode position on terahertz radiation from pentagonal mesas of superconducting Bi$_2$Sr$_2$CaCu$_2$O$_{8+\delta}$, IEEE Trans. THz Sci. Tech. {\bf 5}, 505 (2015). \\
\noindent{[25]}. T. Kashiwagi, K. Sakamoto, H. Kubo, Y. Shibano, T. Enomoto, T. Kitamura, K. Asanuma, T. Yasui, C. Watanabe, K. Nakade,  Y. Saiwai, T. Katsuragawa, M. Tsujimoto, R. Yoshizaki, T. Yamamoto,  H. Minami, R. A. Klemm, and K. Kadowaki, A high-$T_c$ intrinsic Josephson junction emitter tunable from 0.5 to 2.4 terahertz, Appl. Phys. Lett. {\bf 107}, 082601 (2015). \
\noindent{[26]}. T. Kashiwagi, T. Yamamoto, H. Minami, M. Tsujimoto, R. Yoshizaki, K. Delfanazari, T. Kitamura, C. Watanabe, K. Nakade, T. Yasui, K. Asanuma, Y. Saiwai, Y. Shibano, T. Enomoto, H. Kubo, K. Sakamoto, T. Katsuragawa, B. Markovi{\'c}, J. Mirkovi{\'c}, R. A. Klemm, and K. Kadowaki,  Efficient fabrication of intrinsic Josephson junction terahertz oscillators with greatly reduced self-heating effects,  Phys. Rev. Applied {\bf 4}, 054018 (2015). \\
\noindent{[27]}. H. Minami, C. Watanabe, T. Kashiwagi,  T. Yamamoto, K. Kadowaki, and Richard A. Klemm,  0.43 THz emission from high-$T_{\rm c}$ superconducting emitters  at 77 K,  J. Phys.: Condens. Matter {\bf 28}, 025701 (2016).\\
\noindent{[28]}. K. Nakade, T. Kashiwagi, Y. Saiwai, H. Minami, T. Yamamoto, R. A. Klemm, and K. Kadowaki,  Applications using High-$T_c$ terahertz emitters, Sci. Rep. UK {\bf 6}, 23178 (2016). \\
\noindent{[29]}.  C. Watanabe, H. Minami, T. Kitamura, Y. Saiwai, Y. Shibano, T. Katsuragawa, H. Kubo, K. Sakamoto, T. Kashiwagi, R. A. Klemm, and K. Kadowaki,  Electrical Potential Distribution in Terahertz-Emitting Rectangular Mesa Devices of High-$ T_{c}$ Superconducting Bi$_2$Sr$_2$CaCu$_2$O$_{8+\delta}$,   Super. Sci. Tech. {\bf 29}, 065022 (2016).\\
\noindent{[30]}. T. Kashiwagi, H. Kubo, K. Sakamoto, T. Yuasa, Y. Tanabe, C. Watanabe, T. Tanaka, Y. Komori, R. Ota, G. Kuwano, K. Nakamura, T. Katsuragawa, M. Tsujimoto, T. Yamamoto, R. Yoshizaki, H. Minami, K. Kadowaki, and R. A. Klemm, The present status of high-$T_c$ superconducting terhahertz emitters,  Super. Sci. Tech. {\bf 30}, 074008 (2017).\\
\noindent{[31]}. T. Kashiwagi, T. Tanaka, C. Watanabe, H. Kubo, K. Sakamoto, T. Katsuragawa, T. Yuasa, Y. Komori, Y. Tanabe, R. Ota, G. Kuwano, M. Tsujimoto, R. Yoshizaki, H. Minami, T. Yamamoto, R. A. Klemm, and K. Kadowaki, Thermoreflectance microscopy measurements of the Joule heating characteristics of high-$T_c$ superconducting terahertz emitters,  J. Appl. Phys. {\bf 122}, 233902 (2017). \\
\noindent{[32]}. T. Kashiwagi, T. Yuasa, Y. Komori, Y. Tanabe, T. Tanaka, R. Ota, G. Kuwano, K. Nakamura, M. Tsujimoto, H. Minami, T. Yamamoto, R. A. Klemm, and K. Kadowaki,  Improved excitation mode selectivity of high-$T_c$ superconducting terahertz emitters, J. Appl. Phys. {\bf 124}, 033901 (2018). \\
\noindent{[33]}. Y. Shibano, T. Kashiwagi, Y. Komori, K. Sakamoto, Y. Tanabe, T. Yamamoto, H. Minami, R. A. Klemm, and K. Kadowaki,  High-$T_c$ superconducting THz emitters fabricated by wet etching, AIP Advances {\bf 9}, 015116 (2019). \\
\noindent{[34]}. K. Delfanazari, R. A. Klemm, H. J. Joyce, D. A. Ritchie, and K. Kadowaki,  Integrated, portable, tunable, and coherent terahertz sources and sensitive detectors based on layered superconductors,  Proc. IEEE {\bf 108}, 721-734 (2020). \\
\noindent{[35]}. Y. Ono, H. Minami, G. Kuwano, T. Kashiwagi, M. Tsujimoto, K. Kadowaki, and R. A. Klemm,  Superconducting emitter powered at 1.5 THz by an external resonator,  Phys. Rev. Applied {\bf 13}, 064026 (2020).\\
\noindent{[36]}. Y. Saiwai, T. Kashiwagi, K. Nakade, M. Tsujimoto, H. Minami, R. A. Klemm, and K. Kadowaki,  Liquid helium-free high-$T_c$ superconducting terahertz emission system and its applications,  Jpn. J. Appl. Phys. {\bf 59}, 105004 (2020).\\
\noindent{[37]}. M. Zhang, S. Nakagawa, Y. Enomoto, Y. Kuzumi, R. Kikuchi, Y. Yamauchi, T. Hattori, R. A. Klemm, K. Kadowaki, T. Kashiwagi, and K. Delfanazari,  Terahertz source-on-a-chip with decade-long stability using layered superconductor elliptical microcavities,  Phys. Rev. Applied {\bf 24}, 054012 (2025).\\
\noindent{[38]}. M. Tsujimoto, I. Kakeya, T. Kashiwagi, H. Minami, and K. Kadowaki,  Cavity mode identification for coherent terahertz emission from high-$T_c$ superconductors, Opt. Exp. {\bf 24}, (5) 4591 (2016).\\
\noindent{[39]}. R. Kleiner and H. Wang,  Terahertz emission from Bi$_2$Sr$_2$CaCu$_2$O$_{8+x}$  intrinsic Josephson junction stacks, J. Appl. Phys. {\bf 126}, 171101 (2019].\\ 
\noindent{[40]}. Y. Saito, S. Adavhi, R. Matsumoto, M. Nagao, S. Fujita, K. Hayama, K. Terashima, H. Takeya, I. Kakeya, and Y. Takano,   THz emission from a Bi$_2$Sr$_2$CaCu$_2$O$_8$ cross-whisker junction, Appl. Phys. Exp. {\bf 14}, 033003 (2021).\\
\noindent{[41]}. K. J. Kihlstrom, K. C. Reddy, S. Elghazoly, T. E. Sharma, A. E. Koshelev, U. Welp, Y. Hao, R. Divan, M. Tsuijimoto, K. Kadowaki, W.-K. Kwok, and T. M. Benseman,  Powerful terahertz emission from a Bi$_2$Sr$_2$CaCu$_2$O$_{8+\delta}$ mesa operating above 77 K, Phys. Rev. Applied {\bf 19}, 034055 (2023).\\
\noindent{[42]}. M. Miyamoto, R. Kobayashi, G. Kuwano, M. Tsujimoto, and I. Kakeya,  Wide-band frequency modulation of a terahertz intrinsic Josephson junction emitter of a cuprate superconductor, Nat. Photon. {\bf 18}, 266-275 (2024).\\
\noindent{[43]}. K. Delfanazari,  On-chip coherent terahertz emitters with gigahertz modulation, Nat. Photon. {\bf 18}, 214-215 (2024).\\
\noindent{[44]}. S. Gu{\'e}non, M. Gr{\"u}nzweig, B. Gross, J. Yuan, Z. G. Jiang, Y. Y. Zhong, M. Y. Li, A. Iishi, P. H. Wu, T. Hatano, R. G. Mints, E. Goldobin, D. Koelle, H. G. Wang, and R. Kleiner, Interaction of hot spots and terahertz waves in Bi$_2$Sr$_2$CaCu$_2$O$_8$ intrinsic Josephson junction stacks of various geometry, Phys. Rev. B {\bf 82}, 214506 (2010).\\
\noindent{[45]}. A. Elarabi, Y. Yoshioka, M. Tsujimoto, and I. Kakeya, Monolithic superconducting emitter of tunable circularly polarized terahertz radiation, Phys. Rev. Appl. {\bf 8}, 064034 (2017).\\
\noindent{[46]}. E. A. Borodianskyi and V. M. Krasnov, Josephson emission with frequency span 1-11 THz from small  Bi$_2$Sr$_2$CaCu$_2$O$_{8+\delta}$ mesa structures, Nat. Commun. {\bf 8}, 1742 (2017).\\
\noindent{[47]}. L. Ozyuzer, A. E. Koshelev, C. Kurter, N. Gopalsami, Q. Li, M. Tachiki, K. Kadowaki, T. Yamamoto, H. Minami, H. Yamaguchi, T. Tachiki, K. E. Gray, W.-K. Kwok, and U. Welp, Emission of coherent THz radiation from superconductors, Science {\bf 318}, 1291 (2007).\\
\noindent{[48]}. U. Welp, K. Kadowaki, and R. Kleiner, Superconducting emitters of THz radiation, Nat. Photon. {\bf 7}, 702 (2013).\\
\noindent{[49]}. N. Shouk, R. Shouk, S. Bonnough, and R. A. Klemm, Terahertz emission from thin annular and slitted annular Bi2212 microstrip antennas, 2020 Int. Conf. on UK-China Emerging Technologies (UCET) (2020).  doi: 10.1109/UCET51115.2020.9205488\\
\noindent{[50]}. R. A. Klemm, K. Delfanazari, M. Tsujimoto, T. Kashiwagi, T. Kitamura, T. Yamamoto, M. Sawamura, K. Ishida, T. Hattori, and K. Kadowaki, Modeling the electrodynamic cavity mode contributions to the THz emission from triangular Bi$_2$Sr$_2$CaCu$_2$O$_{8+\delta}$ mesas, Physica C {\bf 491}, 30 (2013).\\

\end{document}